\documentclass[a4paper,11pt]{article}
\pdfoutput=1 

\usepackage{jcappub} 

\usepackage[T1]{fontenc} 

\usepackage{graphicx}
\usepackage{subcaption}

\newcommand{\be}{\begin{equation}}
\newcommand{\ee}{\end{equation}}
\newcommand{\bea}{\begin{eqnarray}}
\newcommand{\eea}{\end{eqnarray}}

\title{\boldmath Reexamination of the warm inflation curvature 
perturbations spectrum}

\author[a]{Mar Bastero-Gil,}
\author[b]{Arjun Berera}
\author[b]{and Jaime R. Calder\'on}

\affiliation[a]{Departamento de F\'isica Te\'orica y del Cosmos, Universidad de Granada, Granada-18071, Spain}
\affiliation[b]{School of Physics and Astronomy, University of Edinburgh, Edinburgh, EH9 3FD, United Kingdom}

\emailAdd{mbg@ugr.es}
\emailAdd{ab@ph.ed.ac.uk}
\emailAdd{jaime.calderon@ed.ac.uk}

\abstract{The two approaches to 
compute perturbations in warm inflation are examined. 
It is shown that both 
approaches lead to different expressions for the amplitude of the primordial 
spectrum, with a difference between them of $\Upsilon/(4H)$ at leading order,
where $\Upsilon$ is the dissipation coefficient. In terms of observables, this discrepancy 
can lead to the spectral index differing by up to order $\mathcal{O}(10^{-3})$, which 
is within precision demands for current CMB data. Thus, it is important to resolve this ambiguity to have reliable predictions from
warm inflation.  For this
we prove the extent of this discrepancies by deriving a formula for the 
spectral index and the tensor-to-scalar ratio in each approach. In doing so, 
we find disparities to be more noticeable in a regime where dissipation 
is comparable with the expansion rate, which is a very important regime
from a 
phenomenological point of view.  To determine the extent of
the discrepancy, several cases are examined, including 
quadratic, quartic and hybrid 
potentials with quadratic and $T-$dependent dissipative coefficients.
The origin of the discrepancy is found to be due to the approximation
performed in one of the methods, which underestimates the variation of
the momentum perturbation with expansion.  Once this
is corrected, both approaches are then in agreement.
}

\begin{document}
\maketitle
\flushbottom

\section{Introduction}

It has been around 40 years since the idea of cosmological inflation was 
introduced \cite{Guth1981,LINDE1982389,LINDE1982335,ASinf,Gliner1975,BROUT197878,Brout1979,
STAROBINSKY198099,FANG1980154,kazanas1980dynamics,kolb1980spontaneous}. 
This idea 
solved the problems with the standard Hot Big Bang Cosmology, 
and most importantly, provides 
a mechanism for structure formation through the amplification of 
quantum fluctuations of the scalar field driving inflation. No other 
alternative to inflation has accomplished this much, 
which explains why it is almost 
considered as part of the standard model of Cosmology. In spite of these 
achievements, inflation comes with some problems of its own. For instance, 
the transition between the inflationary and the Hot Big Bang eras requires 
a so-called reheating phase, a very complex and troublesome process where the 
inflaton decays into degrees of freedom from which the known (and unknown) 
particles will ultimately emerge. Another complication in the standard 
inflationary picture is related to the flatness of the scalar field 
potential. Indeed, radiative corrections can spoil the required (lack of) 
steepness of the potentials, so the slow-roll conditions required to 
achieve the necessary 50-60 e-folds of inflation may not be satisfied. 

Against this background, an alternative inflationary scenario, 
warm inflation (WI) \cite{Berera1995}, has proven to be appealing. Within this paradigm the inflaton is allowed to dissipate 
energy into lighter degrees of freedom while inflation is taking place. 
This has far-reaching consequences and advantages when compared to the 
standard cold inflation (CI) scenario. For instance, the
presence of dissipation allows 
for looser slow-roll conditions, so that steeper potentials 
are supported \cite{BASTERO-GIL2009}. Furthermore, a
reheating phase is not needed provided that dissipation is strong enough. 
Thus, a smooth transition between inflation and the Hot Big Bang regime
is possible. 

The warm inflation scenario has also proven to have significant
predictive content in comparison to observational data.
The presence of radiation energy during
inflation results in the fluctuations produced during warm inflation
to be thermal rather than the quantum fluctuations of cold inflation.
Since thermal fluctuations in general are larger than quantum fluctuations,
in order to normalize the scalar spectrum to CMB data, it generally
requires, in a like-for-like comparison of 
the same inflation potential, a lower energy scale
of inflation in warm versus cold inflation.
Moreover the presence of dissipation slows the motion of
the background inflaton field. This means for the same amount of
inflation, the inflaton generally moves less in warm inflation
compared to cold inflation.  The result from both these effects
is to lower the energy scale of inflation in warm inflation.  
For monomial models
such  as $\phi^2$ and $\phi^4$, this was shown in one of the earliest
warm inflation models two decades back \cite{BERERA2000666} (and recently further examined in \cite{bastero2018warm}), and subsequently
it was understood that in general the tensor-to-scalar ratio in
warm inflation for monomial models in suppressed 
relative to cold inflation \cite{BASTERO-GIL2009}.
This correctly predicted the results for the tensor-to-scalar
ratio found in the CMB such as by
WMAP and more decisively by Planck, well before 
they made their observations.

This success of warm inflation is noteworthy since in cold inflation
the monomial models are no longer consistent with data without
reliance on more elaborate model building involving details
about the gravitational interaction of the scalar field.
For decades the monomial models have been the focal point of
cold inflation primarily due to the argument of simplicity.
However the current state of model building required for these models
in cold inflation can no longer rest on that argument. On the
other hand, for warm inflation the model itself is simple, its
only that the calculations are moderately challenging due to
accounting for the effects of interactions and radiation.

In this context, two approaches were developed to compute perturbations in WI. 
On one side, de Oliveira (DO) \cite{DeOliveira2002} derived an analytical expression for the comoving 
curvature perturbation in terms of the field perturbation through suitable 
approximations in the metric and field perturbations equations written in 
the zero-shear gauge. Subsequently, Del Campo \cite{DelCampo2007} built 
upon de Oliveira's work generalising the results by also incorporating in 
the analysis viscous pressure terms. On the other side, and since the 
seminal WI papers, Bastero-Gil, Berera and Ramos 
(BBR) \cite{Berera1995,BASTERO-GIL2009,Ramos2013} derived an expression for 
the amplitude of the primordial power spectrum analogous to the CI scenario. 
As QFT models for the dissipative coefficient were developed, coefficients 
depending on the temperature and the field needed to be considered. 
Naturally, that translated into more complicated perturbation equations. 
Accordingly, numerical studies followed \cite{Bastero-Gil2011} which 
among other things, backed the BBR expression, for a range
of dissipative coefficients.

Although not evident at first glance, the expression found in de Oliveira's 
work does not yield  the  same results as that of Bastero-Gil, Berera and 
Ramos.  This is unexpected and inconvenient, for as experimental 
results become more and more precise, it is crucial to have a consolidated 
program for computing observables in WI.  In this work we explore
the origin of this discrepancy.  To do this, first we rederive the
curvature perturbation power spectrum proposed by both
approaches, de Oliveira and 
Bastero-Gil, Berera and Ramos. From this,
discrepancies between the two approaches are found, even though 
the starting point is a set of equivalent perturbations equations. We then 
pin down the origin of the discrepancies. 
In particular we find the approximations used by DO
underestimates the variation of
the momentum perturbation with expansion. Once this is corrected,
both approaches are in agreement.
We also explore the extent to which this discrepancy  affects
results 
by explicitly computing the spectral index 
and the tensor-to-scalar ratio for several potentials and dissipative 
coefficients using both approaches.  
As an added step, in deriving the BBR comoving curvature perturbation
from a set of gauge invariant equations,
we examine every
term that could potentially introduce corrections both at leading and
next-to-leading order.  No such terms were found, so the original BBR expression is found to be reliable at both orders of approximation.

\section{Background Dynamics}

Warm inflation can be seen as the more general picture in
which the standard cold inflation picture is subsumed.
In warm inflation
the scalar field $\phi$ is allowed to dissipate energy during the 
accelerated expansion inflationary period. The continuous dissipation 
into lighter degrees of 
freedom can be accounted for through a friction-like term $\Upsilon$ in 
the equation of motion of the background scalar field
\begin{equation}\label{eq::bgr}
 \ddot{\phi} + (3H + \Upsilon)\dot{\phi} + V_{\phi} = 0,
\end{equation}
where subscripts denote partial derivatives.  More generally
this equation can be derived from the conservation of the 
energy-momentum tensor, i.e.,
\begin{equation}\label{eq::cons}
	\nabla_a T^{ab} = 0,
\end{equation}
which leads to a set of continuity equations for each component of the cosmological fluid, 
\begin{equation}
	\dot{\rho}_i + 3H(\rho_i + p_i) = \mathcal{Q}_i,
\end{equation}
where $\mathcal{Q}_i$ is known as the source term for the fluid component $i$, such that $\sum_i \mathcal{Q}_i=0$. In particular, in the WI scenario we need to consider in addition to the inflaton field, the radiation as part of the cosmological fluid. In doing so, we get the following equations,
\begin{equation}\label{eq::phid1}
	\dot{\rho_{\phi}}  + 3H(\rho_{\phi} + p_{\phi}) = - \Upsilon(\rho_{\phi} + p_{\phi}),
\end{equation}
\begin{equation}\label{eq::radd1}
	\dot{\rho_{r}}  + 4H\rho_{r} = \Upsilon(\rho_{\phi} + p_{\phi}).
\end{equation}	
The source terms reflect the fact that the origin of the radiation energy density is the energy dissipated by the inflaton field. Furthermore, considering that for a scalar field the energy density and pressure are given by
\begin{equation}\label{eq::rpphi}
	\rho_{\phi} = \frac{1}{2} \dot{\phi}^2 + V(\phi), \hspace{1cm} p_{\phi} = \frac{1}{2} \dot{\phi}^2 - V(\phi),
\end{equation}
we recover then \eqref{eq::bgr} from \eqref{eq::phid1}. 

On the other hand, inflation is conceived as a period of accelerated expansion of the universe, i.e., an era where the scale factor satisfies the condition $\ddot{a} > 0$. From the so-called second Friedmann equation,
\begin{equation}\label{eq::2fr}
	\frac{\ddot{a}}{a} = -\frac{1}{6m_p^2} (\rho + 3p),
\end{equation}
one can see immediately that inflation takes place when $\rho <
-3p$. For a single-field inflationary model this condition is
accomplished when $\dot{\phi}^2 < V(\phi)$. In addition, we also
impose $\ddot{\phi} \ll 3H\dot{\phi}$ so that the kinetic term does
not grow too fast to dominate over the potential term before having
$50-60$ e-folds of expansion. This idea can be generalised readily for
WI. For this, we introduce a dissipative coefficient ratio $Q =
\Upsilon/3H$, 
so the slow-roll approximation yields
\begin{equation}\label{eq::srbf}
	3H(1+Q)\dot{\phi} \simeq -V_{\phi},
\end{equation}
a less restrictive condition than its CI counterpart. The validity of the slow-roll approximation is conveniently checked by looking at the slow-roll parameters, which can be derived by taking the derivative of both sides of the equation above with respect to the number of e-folds, getting
\begin{equation}\label{eq::srdph}
	\frac{d \ln \dot{\phi}}{dN}  = \frac{d\ln V_{\phi}}{dN} - \frac{d \ln H}{dN} - \frac{d \ln (1+Q)}{dN} \,,
\end{equation}
where the modululus of each term should be small compared to 1 for the approximation to be valid. In this way, the set of slow-roll parameters is given by
\begin{equation}\label{srw0}
	\epsilon  = -\frac{d \ln H}{dN}, \hspace{1cm} \eta = -\frac{d \ln V_{\phi}}{dN}, \hspace{1cm} \theta = \frac{d \ln (1+Q)}{dN}.
\end{equation}
The first two parameters are defined analogously to the CI scenario \citep{Liddle1994}, so that during  slow-roll WI they can be written like 
\begin{equation}\label{srwi1}
\epsilon \simeq \frac{\epsilon_{\phi}}{1+Q} = \frac{m_p^2}{2(1+Q)} \left(\frac{V_{\phi}}{V}\right)^2 \ll 1, \hspace{2cm} |\eta| \simeq \frac{|\eta_{\phi}|}{1+Q} = \frac{m_p^2}{1+Q} \left| \frac{V_{\phi \phi}}{V} \right| \ll 1.
\end{equation}
The specific form of the parameter $\theta$, even during the
slow-roll regime, depends on the specific form of
$\Upsilon$. General dissipative coefficients depending on both $T$ and
$\phi$,  $\Upsilon \propto T^c/\phi^m$, are
particularly interesting from a phenomenological and theoretical point
of view \cite{Bastero-Gil2011,Ramos2013}. In Appendix \ref{s::tdc} the
general expression for $\theta$ is given in terms of the other slow-roll
parameters, Eq. \eqref{dlambda}. 
That expression has also introduced a slow-roll parameter 
describing the change on the scalar field during inflation, 
\begin{equation}\label{srwi2}
	\sigma = -\frac{d \ln \phi}{dN} \simeq \frac{m_p^2}{1+Q} \frac{V_{\phi}/\phi}{V} \,.
\end{equation}

Finally, the slow-roll approximation should be also applied to 
the radiation energy density \eqref{eq::radd1}, which gives
\begin{equation}\label{eq::srraden}
	4H \rho_r \simeq \Upsilon \dot{\phi}^2 \,.
\end{equation}
We go under the assumption that $\rho_r \ll \rho_{\phi} \simeq V$, otherwise inflation would end since the inflaton would not be driving the background dynamics of the expansion. This is consistent with Eq. \eqref{eq::srraden}, which can be written as:
\be
\frac{\rho_r}{V} \simeq \frac{Q}{2 (1+Q)} \epsilon \,,
\ee
showing that indeed during slow-roll WI with $\epsilon <1$ the
radiation energy density is subdominant. 

\section{Cosmological Perturbations and WI}

Cosmological perturbations are closely tied to observables. Indeed,
fluctuations during the inflationary era are believed to have left
fingerprints that persist to this day in the form of anisotropies in
the Cosmic Microwave Background. In this way, when we survey the CMB
we have access to information about the scale dependence of primordial 
perturbations (spectral index $n_s$), and the relation between scalar
and tensor perturbations (tensor-to-scalar ratio $r$). Thus, the
predictions of a specific inflationary model are given in terms of
these magnitudes, which characterize the primordial power spectrum of 
of the gauge invariant quantity known as the comoving curvature perturbation
$\mathcal{R}$, 
\be 
\Delta_{\cal R}^2(k)= \frac{k^3}{2 \pi^2} |{\cal R}_k|^2 \,,
\ee
where $k$ is the comoving wavenumber. The amplitude of modes crossing
the horizon around 60-50 e-folds\footnote{The value of the number of efolds at which the largest observable scale crossed the horizon depends on details of the reheating process, i.e., how the universe becomes radiation dominated after inflation \cite{Adshead:2010mc}.} before the
end of inflation should match that observed by the Planck mission 
$A_0= 2.9 \times 10^{-9}$ \cite{Ade:2015xua}. 
Consequently, as the precision of cosmological and CMB
surveys increases, so does the need to perform more careful
calculations of the curvature perturbation.

In this context, the WI paradigm presents an interesting case of
study. Unlike the standard inflationary picture, where quantum
fluctuations of the scalar field are the origin of perturbations, WI
posits that if the inflaton interacted with a heat bath, the
corresponding thermal fluctuations are the primary source of
perturbations. This has important consequences, as it has been
found that some potentials otherwise discarded by CI can be rendered
consistent with observations \cite{Rangarajan:2018tte}. 

In this section, 
the basic calculation of the
comoving curvature perturbation ${\cal R}_k$ during WI is 
reviewed. For this, 
the perturbed Friedmann-Lema\^itre-Robertson-Walker (FLRW) metric is,
\begin{equation}\label{eq::met}
ds^2 = -(1+2\alpha)dt^2 -2a \partial_i \beta dx^i dt + a^2 \left[\delta_{ij}(1+2\varphi) + 2 \partial_i \partial_j \gamma \right] dx^i dx^j \;,
\end{equation}
together with the metric-related variables:
\begin{equation}
\chi = a(\beta + a \dot{\gamma}),
\end{equation}
\begin{equation}
\kappa = 3(H \alpha - \dot{\varphi}) + \partial_k \partial^k \chi,
\end{equation}
where $\chi$ is known as the shear and $-\kappa$ is the perturbed expansion. These variables are related with perturbed quantities through energy and momentum constraint equations, given by:
\begin{equation}\label{eq::einper1}
	H \kappa - \frac{k^2}{a^2} \varphi = -\frac{\delta \rho_T}{2 m_p^2},
\end{equation}
\begin{equation}\label{eq::einper2}
	-\dot{\varphi} + H \alpha =- \frac{\Psi_T}{2m_p^2},
\end{equation}
where $\Psi_T$ and  $\delta \rho_T$ denote the total momentum and
energy density perturbations, respectively. 

The equation of motion of the perturbed scalar field is given by
\begin{equation}\label{pert1}
	{\delta \ddot \phi} + (3H + \Upsilon){\delta \dot \phi} + \left(\frac{k^2}{a^2} + V_{\phi \phi} \right) \delta \phi  =  - \delta \Upsilon \dot{\phi} + \dot{\phi} (\kappa + \dot{\alpha}) + (2\ddot{\phi} + (3H + \Upsilon)\dot{\phi}) \alpha,
\end{equation}
i.e., the perturbed version of \eqref{eq::bgr}. On the other hand, from the conservation of the energy-momentum tensor, the equations of motion for the perturbed radiation energy density $\delta \rho_r$ and momentum $\Psi_r$ in the absence of shear viscous pressure read:
\begin{eqnarray}
	{\delta\dot \rho}_r + 4H\delta\rho_r & = & \frac{k^2}{a^2} \Psi_r + \dot{\rho}_r \alpha + \frac{4}{3}\rho_r\kappa + \delta \mathcal{Q}_r \label{pert2}, \\
	\dot{\Psi}_r + 3H \Psi_r & = & -\frac{1}{3}\delta \rho_r - \frac{4}{3} \rho_r \alpha + \mathbf{J}_r  \label{pert3},
\end{eqnarray}
where $\Psi_{\phi} = -\dot{\phi} \delta \phi$, and $\textbf{J}_{r} = \Upsilon \Psi_{\phi}$ is the momentum source.  We have also used the equation of state for the radiation degrees of freedom  $\delta p_r = \delta \rho_r/3$. Regarding the field-related equations, all the information contained in the equations of motion of $\Psi_{\phi}$ and $\delta \rho_{\phi}$ is already encompassed in \eqref{pert1}.

Next, the evolution equations are rewritten in terms of gauge invariant
quantities.  In this way it becomes easier to compare magnitudes in
different gauges. To do this, following \cite{Hwang1991}, we introduce gauge invariant scalar perturbations,  
\begin{equation}
    \delta f^{GI} = \delta f - \frac{\dot{f}}{H} \varphi,
\end{equation}
where $f$ is a scalar background magnitude, such as the energy
density, pressure or the scalar field. On the other hand, one 
can define a gauge invariant momentum perturbation such that
\begin{equation}
    \Psi_{\alpha}^{GI} = \Psi_{\alpha} + \frac{\rho_{\alpha}+ p_{\alpha}}{H} \varphi.
\end{equation}
Finally, the following gauge invariant metric perturbations are introduced:
\begin{equation}
    \mathcal{A} = \alpha - \frac{\dot{\varphi}}{H} - \epsilon \varphi,
\label{eq:calA}
\end{equation}
\begin{equation}
    \Phi = \varphi - H \chi.
\end{equation}

At this point, it is straightforward to write the gauge invariant version of the equations governing the evolution of perturbations. The resulting equations are:
\begin{eqnarray}\label{eq::dph}
	\ddot{\delta \phi}^{GI} + (3H + \Upsilon) \dot{\delta \phi}^{GI} + \left(\frac{k^2}{a^2} + V_{\phi \phi} \right) \delta \phi^{GI} & = & - \dot{\phi}\delta \Upsilon^{GI} + \Upsilon \dot{\phi} \mathcal{A} + \dot{\phi} \dot{\mathcal{A}} + 2(\ddot{\phi} + 3H \dot{\phi})\mathcal{A}  \nonumber \\
	& & - \frac{k^2}{a^2} \dot{\phi} \frac{\Phi}{H}, \label{eq::dp}
\end{eqnarray}
\begin{equation}\label{eq::dr} 
\delta \dot{\rho_r}^{GI} + 4H \delta \rho_r^{GI}   =   \frac{k^2}{a^2} \Psi_r^{GI} + \delta \mathcal{Q}_r^{GI} + \dot{\rho_r} \mathcal{A} - 2 H \frac{\rho_r}{\rho_T} \delta \rho^{GI}_T, 
\end{equation}
\begin{equation}\label{eq::dq}
\dot{\Psi}_r^{GI} + 3H \Psi_r^{GI}   =   -\frac{1}{3} \delta \rho_{r}^{GI}  + \Upsilon \Psi_{\phi}^{GI} - \frac{4}{3} \rho_r\mathcal{A} \;,
\end{equation}
and the Einstein equations curbing energy density and momentum 
perturbations  can be written as
\begin{equation}\label{ein1}
    \frac{k^2}{a^2 H^2}\Phi = 3 \mathcal{A} + \frac{3}{2} \frac{\delta \rho^{GI}_T}{\rho_T},
\end{equation}
\begin{equation}\label{ein2}
    \mathcal{A} = \epsilon \mathcal{R},
\end{equation}
where $\mathcal{R}$ is known as the comoving curvature perturbation, defined by
\begin{equation}\label{eq::rmgi}
    \mathcal{R} = -\frac{H}{\rho+p}\Psi_T^{GI} = -\frac{1}{2m_p^2}\frac{1}{\epsilon H} (\Psi_{\phi}^{GI} + \Psi_r^{GI}).
\end{equation}

The comoving curvature perturbation is a gauge invariant magnitude itself that measures the spatial curvature of comoving hypersurfaces. Furthermore, for superhorizon modes it also measures the curvature of constant-density hypersurfaces \cite{Baumann2009}. In general, for a multifluid model with different components $\alpha$, like WI, it can be written as:
\be \label{reqs}
\mathcal{R} \simeq \sum_\alpha \frac{\rho_\alpha + p_\alpha}{\rho +p } \mathcal{R}_{\alpha} \,,
\ee
where
\begin{equation}\label{rcomp}
	\mathcal{R}_{\alpha} = -\frac{H}{\rho_{\alpha} + p_{\alpha}} \Psi_{\alpha}^{GI}\,.
\end{equation}

\section{Computing perturbations in WI}

As mentioned previously, our task is to compute the comoving curvature
perturbation. The usual plan of action consists in solving the
equations of motion of the perturbations, either
Eqs. \eqref{pert1}-\eqref{pert3} in a particular gauge, or
Eqs. \eqref{eq::dph}-\eqref{eq::dq}. Naturally, one way to do it is
through numerical simulations \cite{Bastero-Gil2011}, although analytical implementations are the focus of this work. We will address the main
implementations, one proposed originally by de Oliveira in
\cite{DeOliveira2002}, and the second one by authors in
Refs. \cite{Hall2004, Graham:2009bf, Bastero-Gil2011}. In both cases
the goal is to get an expression for the comoving curvature
perturbation in terms of the scalar field perturbation and background
quantities. As it can be inferred from the equations governing the
evolution of perturbations, this is not a simple task, 
and several approximations are in order. It will be seen that there are
small discrepancies that lead to different predictions for observables
between the two approaches. The goal is to clarify the origin of those
discrepancies, and identify the regime where such discrepancies
will have noticeable effects.   

\subsection{de Oliveira (DO) implementation}

The first analytical implementation we will cover was introduced by de
Oliveira almost 20 years ago in \cite{DeOliveira2002} for monomial
dissipative coefficients in the absense of anisotropic stress, and
later generalised by Del Campo in \cite{DelCampo2007}, who did
consider such terms. For our purposes, it will be enough to focus on
the case treated in \cite{DeOliveira2002}. He used the zero-shear or Newtonian gauge, so the metric \eqref{eq::met} reduces to 
\begin{equation}
	ds^2 = -(1+2\alpha)dt^2 +a^2 (1+2\varphi) \delta_{ij} dx^i dx^j.
\end{equation}
Since no anisotropic stress is considered, Einstein equations imply
$\alpha = -\varphi$. Next, this program set to use a slow-roll--like
approximation in the equations of the perturbations in order to find
an analytical expression for the metric perturbation, and, in doing
so, to find a way to compute observables in the context of COBE
normalization. 
 
Considering the assumptions mentioned above, the set of equations
presented in \cite{DeOliveira2002} is completely equivalent to those
used by other approaches. However, instead of working with the momentum
perturbation (a point that will be key in our analysis), they worked
with a velocity field defined by\footnote{In current conventions the 
comoving wavenumber is not included in the definition of the 
velocity field. However, we have decided to keep this factor in 
order to reproduce the set of equations in the same way as in 
\cite{DeOliveira2002}. Needless to say, both definitions lead to the same
  conclusions.} 
\begin{equation}
	v = \frac{k}{a} \frac{\Psi_r}{\rho_r + p_r},
\end{equation}
which upon replacement in \eqref{pert2} and \eqref{pert3} yields the equations:
\begin{equation}\label{dorr}
	{\delta \dot \rho_r} + 4H \delta \rho_r = \frac{4}{3} \frac{k}{a} \rho_r v + \dot{\rho_r} \alpha + \frac{4}{3} \rho_r \kappa + \delta \mathcal{Q}_r,
\end{equation}
\begin{equation}\label{veom}
	\dot{v} + \frac{\Upsilon \dot{\phi}^2}{\rho_r} v + \frac{k}{a} \left( \alpha + \frac{\delta \rho_r}{4 \rho_r} + \frac{3}{4} \frac{\Upsilon \dot{\phi}}{\rho_r} \delta \phi \right)=0.
\end{equation}
Since only monomial dissipative coefficients were considered, the source perturbation reduces to $\delta \mathcal{Q}_r = 2 \Upsilon \dot{\phi} ({\delta \dot \phi} - \dot{\phi} \alpha) + \Upsilon_{\phi} \dot{\phi}^2 \delta \phi$. Next, by means of the slow-roll approximation, we neglect the higher time derivatives of the perturbations in the equations above. Consequently,
\begin{equation}
	v \simeq - \frac{k}{4aH} \left(\alpha + \frac{\delta \rho_r}{4 \rho_r} + \frac{3}{4} \frac{\Upsilon \dot{\phi}}{\rho_r} \delta \phi \right),
\end{equation}
where the second term inside the brackets can be computed through a
similar approximation in \eqref{dorr}. Finally, the Einstein
equation \eqref{eq::einper2} in the absence of shear pressure, 
 under the slow-roll approximation reads  
\begin{equation}\label{do0}
	\alpha \simeq \frac{1}{2m_p^2} \frac{\dot{\phi}}{H} \left(1 + \frac{\Upsilon}{4H} + \frac{\Upsilon_{\phi}\dot{\phi}}{48H^2} \right) \delta \phi.
\end{equation}
As we are only interested in super-horizon modes, terms proportional
to $k/aH$ are considered vanishingly small, similarly to terms
proportional to $\dot{\delta \phi}/\dot{\phi}$. 
Finally, keeping the leading term in the comoving curvature perturbation formula \eqref{ein2} in the Newtonian gauge yields 
\begin{equation}
	\mathcal{R} \simeq \frac{\alpha}{\epsilon},
\end{equation}
with a curvature power spectrum given by
\begin{equation}
	\Delta_{\mathcal{R}}^2 = \left[\frac{1}{2m_p^2} \frac{\dot{\phi}}{\epsilon H} \left(1 + \frac{\Upsilon}{4H} + \frac{\Upsilon_{\phi}\dot{\phi}}{48H^2} \right)\right]^2 \Delta^2_{\delta \phi}.\\
\end{equation}

This was the result found by de Oliveira. For the purpose of considering dissipative coefficients of the form $\Upsilon \propto T^c/\phi^m$, we can follow a similar recipe. In fact, we only need to notice that
\begin{equation}\label{uptm}
    \frac{\delta \Upsilon}{\Upsilon} = c\frac{\delta T}{T} - m \frac{\delta \phi}{\phi},
\end{equation}
where the temperature and radiation energy perturbations are related by
\begin{equation}
    \frac{\delta \rho_r}{\rho_r} = 4 \frac{\delta T}{T}.
\end{equation}
In that way, we arrive to the following expression
\begin{equation}
    \alpha = \frac{1}{2m_p^2} \frac{\dot{\phi}}{H} \left(1+\frac{\Upsilon}{4H} + \frac{m}{c-4} \frac{\Upsilon \dot{\phi}}{12 H^2 \phi}\right)\delta \phi,
\end{equation}
which can be easily seen to agree with \eqref{do0} for the $c=0$ case. We can express this equation in terms of the dissipative ratio and the slow-roll parameters, such that
\begin{equation}
    \alpha = \frac{1}{2m_p^2} \frac{\dot{\phi}}{H} \left(1 + Q\left(\frac{3}{4} - \frac{m}{4(c-4)} \sigma\right)\right) \delta \phi.
\end{equation}
Then, neglecting once again the highest time derivative, the comoving curvature perturbation is given by
\begin{equation}
    \mathcal{R} \simeq \frac{1}{2m_p^2} \frac{\dot{\phi}}{\epsilon H} \left(1 + Q\left(\frac{3}{4} - \frac{m}{4(c-4)} \sigma\right)\right)\delta \phi,
\end{equation}
and, consequently, its power spectrum reads
\begin{equation}\label{DRai}
    \Delta_{\mathcal{R}}^2 = \left[\frac{1}{2m_p^2} \frac{\dot{\phi}}{\epsilon H} \left\{1 + Q\left(\frac{3}{4} - \frac{m}{4(c-4)} \sigma\right)\right\} \right]^2 \Delta_{\delta\phi}^2.
\end{equation}

Finally, $\Delta_{\delta\phi}^2$ remains to be determined. There
are several articles reviewing this; in particular \cite{Ramos2013},
which deals with how to compute the inflaton power spectrum considering quantum and thermal contributions concomitantly. Explicit comparisons were made using this result. 

\subsection{Bastero-Gil, Berera, Ramos (BBR) implementation}\label{BBR}

In order to derive the BBR result, a similar program to
the previous section will be followed. 
For the sake of consistency and briefness, we
will only work at zeroth order in the slow-roll parameters, leaving an
extension to linear order for Appendix \ref{s::correctionsR}.
The equations governing the evolution of gauge invariant quantities
will be used,
as their connection to the comoving curvature perturbation is more
neat. 

First, the radiation momentum perturbation is computed. 
For this, we refer to its equation of motion \eqref{eq::dq}. 
In the same fashion as de Oliveira, the higher time derivative is
neglected, which yields
\begin{equation}
	\Psi^{GI}_r \simeq Q \Psi_{\phi}^{GI} - \frac{1}{9H} (\delta \rho_r^{GI} + 4 \rho_r \mathcal{A}).
\end{equation}
As it will be discussed in the appendices, the terms inside the parenthesis introduce corrections to the curvature perturbation at a higher order, thus, at zeroth order the momentum perturbation follows the simple expression
\begin{equation}
	\Psi_r^{GI} \simeq Q \Psi_{\phi}^{GI}.
\end{equation}
Then, \eqref{rcomp} implies that the radiation contribution to 
the curvature perturbation is given by
\begin{equation}
	\mathcal{R}_r = -\frac{H}{\rho_r + p_r} \Psi_r^{GI} \simeq -\frac{3}{4} \frac{H}{\rho_r} Q\Psi_{\phi}^{GI} \simeq  - \frac{H}{\dot{\phi}^2} \Psi_{\phi}^{GI}  =  \mathcal{R}_{\phi}.
\end{equation}
Hence, by \eqref{reqs} and the equation above we conclude that 
\begin{equation} \label{rrphirr}
	\mathcal{R} \simeq \mathcal{R}_{\phi} \simeq \mathcal{R}_r
\end{equation}
which leads to the following curvature power spectrum
\begin{equation}\label{bbr1}
	\Delta_{\mathcal{R}}^2 \simeq \Delta_{\mathcal{R}_{\phi}}^2 = \left(\frac{H}{\dot{\phi}}\right)^2 \Delta^2_{\delta\phi}.
\end{equation}
This result has been checked numerically for several dissipative coefficients and potentials in \cite{Bastero-Gil2011}. It is worth noticing 
that \eqref{bbr1} has the same functional form as the CI expression, which is why a similar formula was already used on seminal papers about WI, 
like in \cite{BERERA2000666,berera1995thermally}.
In addition, \eqref{rrphirr} ensures that there are no isocurvature perturbations during warm inflation, and therefore $\dot {\cal R} = 0$ once the perturbation becomes superhorizon.  


\subsection{Discrepancies}\label{ss:dis}

The processes outlined in the previous sections lead to different results, even at
a leading order approximation. In order to understand the origin of
those discrepancies, we write de Oliveira's result as 
\begin{equation}\label{do}
	(\Delta_{\mathcal{R}}^2)^{DO} \simeq \left[\frac{1}{2m_p^2} \frac{\dot{\phi}}{\epsilon H} \left\{1 + \frac{3}{4}Q\right\}\right]^2 \Delta_{\delta\phi}^2,
\end{equation}
where we have dismissed the $\sigma-$term, which introduces corrections to the spectral index and the tensor-to-scalar ratio at second order in the slow-roll parameters. On the other hand, for the sake of comparison, we rewrite \eqref{bbr1} as
\begin{equation}\label{bbr}
	(\Delta_{\mathcal{R}}^2)^{BBR} = \left[\frac{1}{2m_p^2} \frac{\dot{\phi}}{\epsilon H} \{1 + Q\}\right]^2 \Delta_{\delta\phi}^2,
\end{equation}
where we have used the slow-roll approximation \eqref{eq::srbf} together with the definition of the parameter $\epsilon$, \eqref{srwi1}.
Even though these expressions are quite similar, the
discrepancies between the procedures manifest as a difference of
$Q/4$ in the dissipative term. 
This is rather unexpected, since both results were obtained
from equivalent sets of equations and through similar kind of
approximations. Thus, it is natural to infer that somehow, different approximations were made in each case. To go deeper into this, it
is useful to analyse the process followed by de Oliveira. As stated
previously, instead of using the momentum perturbation, he used a
velocity field defined by  
\begin{equation}
	v = \frac{3}{4} \frac{k}{a} \frac{\Psi_r}{\rho_r},
\end{equation}
with a time derivative given by
\begin{equation}
	\dot{v} = \frac{3}{4} \frac{k}{a \rho_r} \left(\dot{\Psi}_r - \frac{\dot{\rho_r}}{\rho_r}\Psi_r -H \Psi_r  \right).
\end{equation}

When one uses the slow-roll approximation in the equation of motion \eqref{veom} of the velocity field, the three terms on the RHS of the equation above are being neglected. However, only the first two involve time derivatives, whereas the last term is actually non-negligible. We check this through numerical simulations in Appendix \ref{s::aplt}. Then, in other words, in neglecting the time derivative of the velocity field we are overlooking part of the dilution of the radiation momentum perturbation expressed in its equation of motion \eqref{pert3} or \eqref{eq::dq}. For this reason the velocity field and the momentum perturbation dilute at different rates during expansion, and more importantly, that is why \eqref{do} is missing a difference of $Q/4$.

The BBR result \eqref{bbr} could be recovered working instead with the
{\it covariant} velocity perturbations \cite{Malik2003}, which for a fluid component is given by
\begin{equation}
	\Psi_{\alpha} = (\rho_{\alpha} + p_{\alpha})V_{\alpha} = \frac{a}{k}(\rho_{\alpha} + p_{\alpha})(v_{\alpha}+k \beta).
\end{equation}
In particular, for the radiation degrees of freedom, the momentum perturbation follows the relation
\begin{equation}
\Psi_r = \frac{4}{3}\rho_r V_r= \frac{4}{3}\frac{a}{k}\rho_r(v + k\beta).
\end{equation}
In consequence, the comoving curvature perturbation can be written as
\begin{equation}
	\mathcal{R} = -\varphi  - H(V_{\phi}+V_r).
\end{equation}
We note that the covariant velocity dilutes at the same rate as the 
momentum perturbation, so if the slow-roll approximation is applied to 
its equation of 
motion it will easily recover the BBR result. 

\subsection{Observables}
The differences presented in the curvature power spectrum will lead to 
different expressions of the spectral index and the tensor-to-scalar ratio. 
In order to show the differences independently of the 
form of $\Delta_{\delta\phi}^2$, each curvature power spectrum is
written as
\begin{equation}
	\Delta^2_{\mathcal{R}i} = \mathcal{K}_i \left(\frac{\dot{\phi}}{2 m_P^2 \epsilon H}\right)^2 \Delta_{\delta\phi}^2,
\end{equation}
where $\mathcal{K}_i$ is given by the square of the terms inside the braces 
in equations \eqref{do} and \eqref{bbr}. Then, the spectral index is
\begin{equation}\label{nsi}
	n_{s/i} - 1 = \frac{d \ln \Delta^2_{\mathcal{R}i}}{dN} = \tilde{n} + \frac{d \ln \mathcal{K}_i}{dN},
\end{equation}
where the last term on the RHS is the only one that depends on the approach. 
In this way, using the appropriate slow-roll equations of motion for the 
case of interest and working at linear order on the slow-roll parameters gives 
\begin{eqnarray}
    n_{s/DO} & = & 1 + \tilde{n} + \frac{6(1+Q)}{4+3Q} \theta + \mathcal{O}(\epsilon^2,\eta^2,\theta^2),\label{nsdo}\\
	n_{s/BBR} & = & 1 + \tilde{n} + 2 \theta.\label{nsbbr}
\end{eqnarray}
It is worth noticing that the two expressions agree in the limit of strong dissipation. They will also agree in the very weak dissipative regime, with $Q \ll 1$, given that $\theta$ is proportional itself to $Q$ (see \eqref{dlambda}).  
However, that is not the case for weak dissipation with $Q \simeq \mathcal{O}(1)$, since for de Oliveira's approach the proportionality factor of $\theta$ is roughly $3/2$, as oppose to the BBR approach, where the analogous factor is $2$ for the entire dissipation regime. As it will be seen in the examples below, this can lead to a difference in the predicted spectral index up to order $\mathcal{O}(10^{-3})$, which is within the sensitivity of Planck results.

Regarding the tensor-to-scalar ratio, it will prove easier to work with the inverse of this magnitude such that
\begin{equation}
    \frac{1}{r} = \frac{\Delta^2_s}{\Delta^2_T},
\end{equation}
where $\Delta^2_s$ denotes the power spectrum of scalar perturbations, or in this case, the curvature perturbation power spectrum. On the other hand, $\Delta^2_T$ represents the power spectrum for tensor perturbations, given by 
\begin{equation}
    \Delta^2_T = \frac{2}{\pi^2}\frac{H^2}{m_p^2}.
\end{equation}
Therefore, the (inverse) tensor-to-scalar ratio for each case is 
\begin{eqnarray}
    \frac{1}{r_{DO}} & = & \frac{\pi^2}{4 H^2 \epsilon_{\phi}}\left\{1+ \frac{3}{4}Q \right\}^2 \Delta^2_{\delta\phi} \,,\\
    \frac{1}{r_{BBR}} & = & \frac{\pi^2}{4 H^2 \epsilon_{\phi}} \left\{1+Q\right\}^2 \Delta^2_{\delta\phi} \,. 
\end{eqnarray}
Contrary to the scalar spectral index, the predictions in both approaches will differ only in the strong dissipative regime, with $r_{DO}$ slightly overestimating the ratio.  

\subsection{Examples}

In this section some examples are presented of the spectral index and the
tensor-to-scalar ratio predictions given by each approach. It is worth
mentioning that this is only a comparative exercise, so we will not
take into account if one result is close or far from the experimental
values, i.e.,  we will only pay attention to the differences between
the two approaches that in principle should lead to the same
results. Regarding the computational aspects, in each case
the predictions are computed considering that the perturbation modes
freeze out $60$ e-folds before the end of inflation, fixing the
parameters to be consistent with  $\Delta^2_\mathcal{R}\simeq
2.5\times10^{-9}$. On the other hand, so far we have kept our
discussion independent of the form of
$\Delta_{\delta\phi}^2$. However, in order to see explicitly the
differences in the predictions, we recur to the result found in
\cite{Ramos2013}: 
\begin{equation} \label{Deltaphi}
	\Delta_{\delta \phi}^2 = \left(\frac{H}{ 2\pi}\right)^2 \left[1+2n_* + \frac{T}{H}\frac{12 Q 8^Q \ [\Gamma(3/2+3Q/2)]^3}{(1+3Q)\Gamma(1+3Q/2)\Gamma(5/2+3Q)}\right] \,,
\end{equation}
where $n_*$ denotes the statistical distribution of the inflaton at horizon crossing, and all variables $H$, $T$ and $Q$ are evaluated at horizon crossing. Using the properties of the $\Gamma$ functions, this expression can be very well approximated by:
\be \label{Deltaphi2}
\Delta_{\delta \phi}^2 = \left(\frac{H}{2 \pi}\right)^2 \left[1+2n_* + \frac{T}{H}\frac{ 2 \pi \sqrt{3} Q}{\sqrt{3 + 4 \pi Q}}\right]\,,
\ee
which is the expression used in recent analyses of warm inflation \cite{Bartrum2014,Bastero-Gil:2016qru, Benetti:2016jhf,Bastero-Gil:2017wwl,Arya:2018sgw,Arya:2017zlb}. And in the limit where the last term within the squared brackets in \eqref{Deltaphi2} dominates independently of $n_*$,  and $Q > 1$, one also recovers the approximated expression derived in \cite{taylor2000perturbation}
\be   
\Delta_{\delta \phi}^2 \sim  \frac{H^2}{2 \pi^2}  
 \frac{T}{H} \sqrt{3 Q} \,.
\ee

\begin{figure}[htp!]
  \centering
  \begin{subfigure}[b]{0.47\textwidth}
    \includegraphics[width=\textwidth]{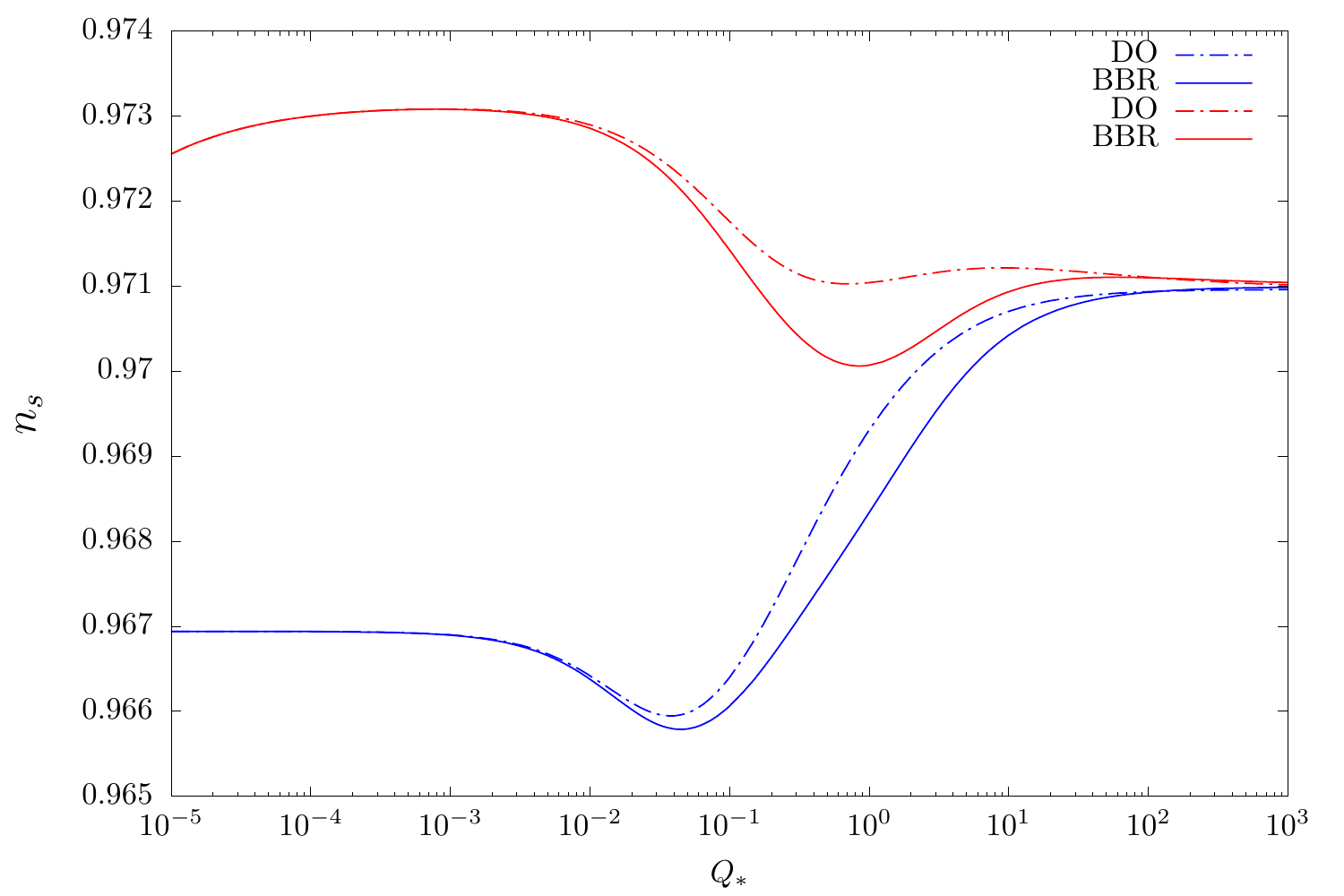}
    \caption{$\Upsilon \propto \phi^2$}
  \end{subfigure}
~
  \begin{subfigure}[b]{0.47\textwidth}
    \includegraphics[width=\textwidth]{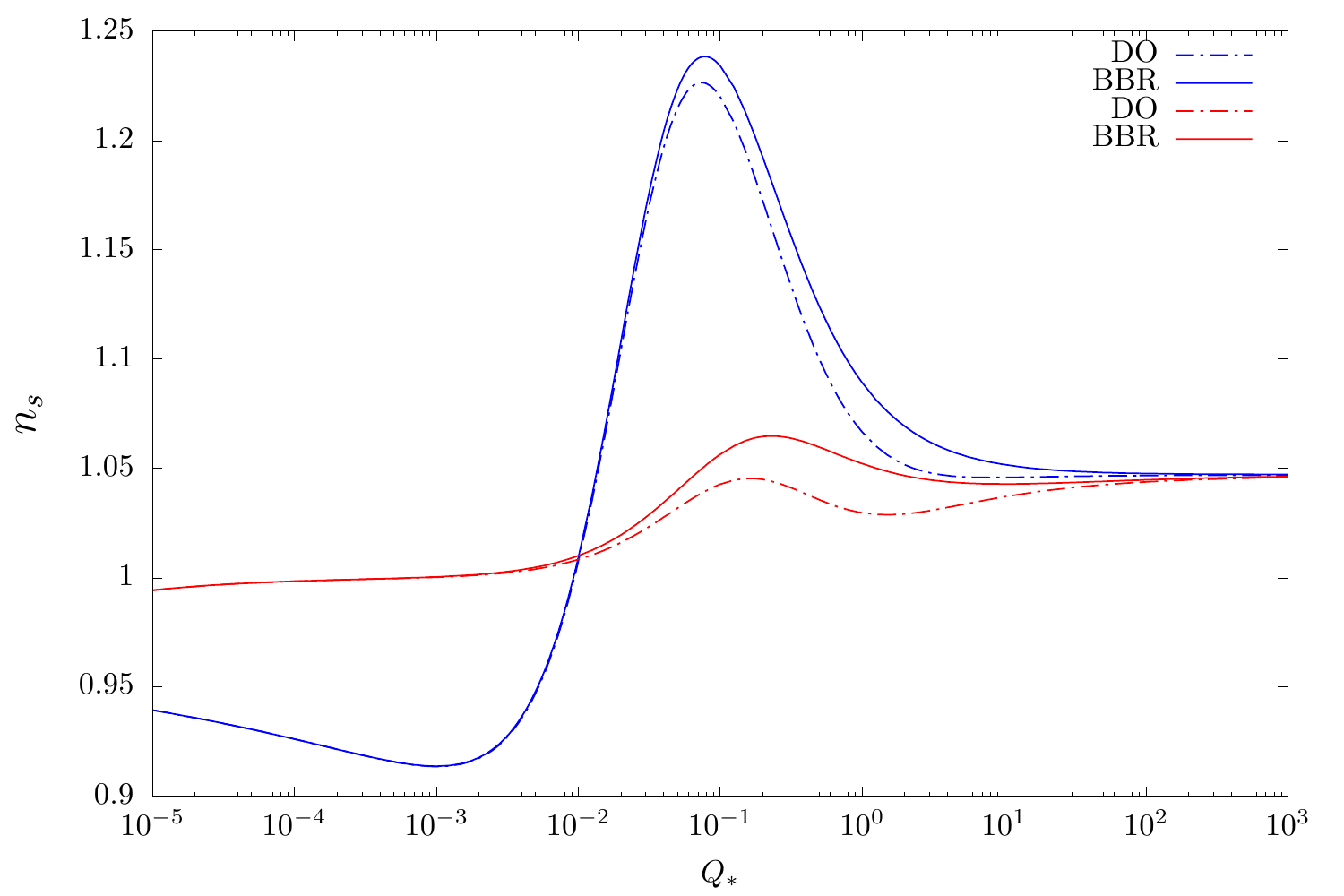}
    \caption{$\Upsilon \propto T^3/\phi^2$}
  \end{subfigure}\\
  \begin{subfigure}[b]{0.47\textwidth}
    \includegraphics[width=\textwidth]{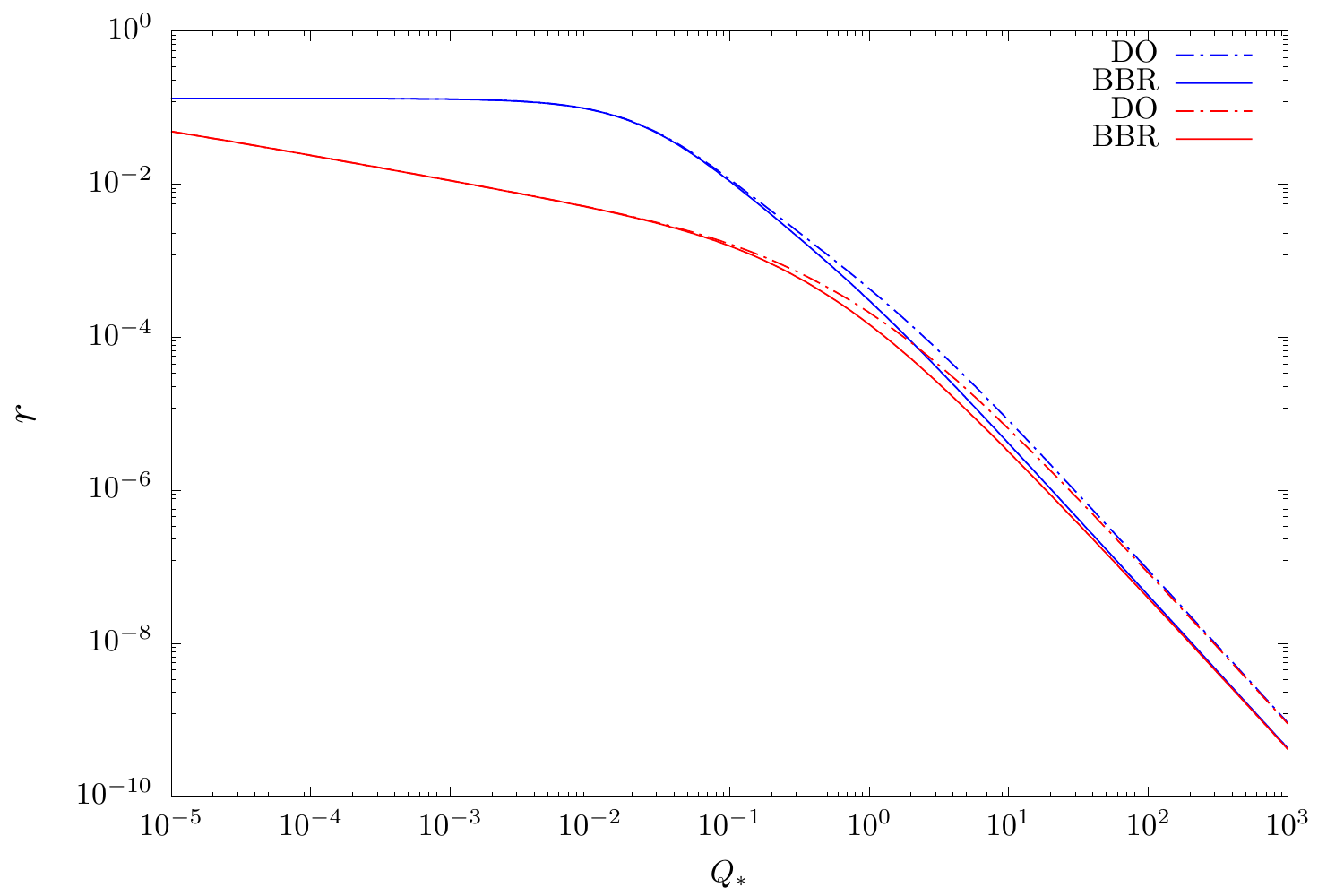}
    \caption{$\Upsilon \propto \phi^2$}
  \end{subfigure}
~
  \begin{subfigure}[b]{0.47\textwidth}
    \includegraphics[width=\textwidth]{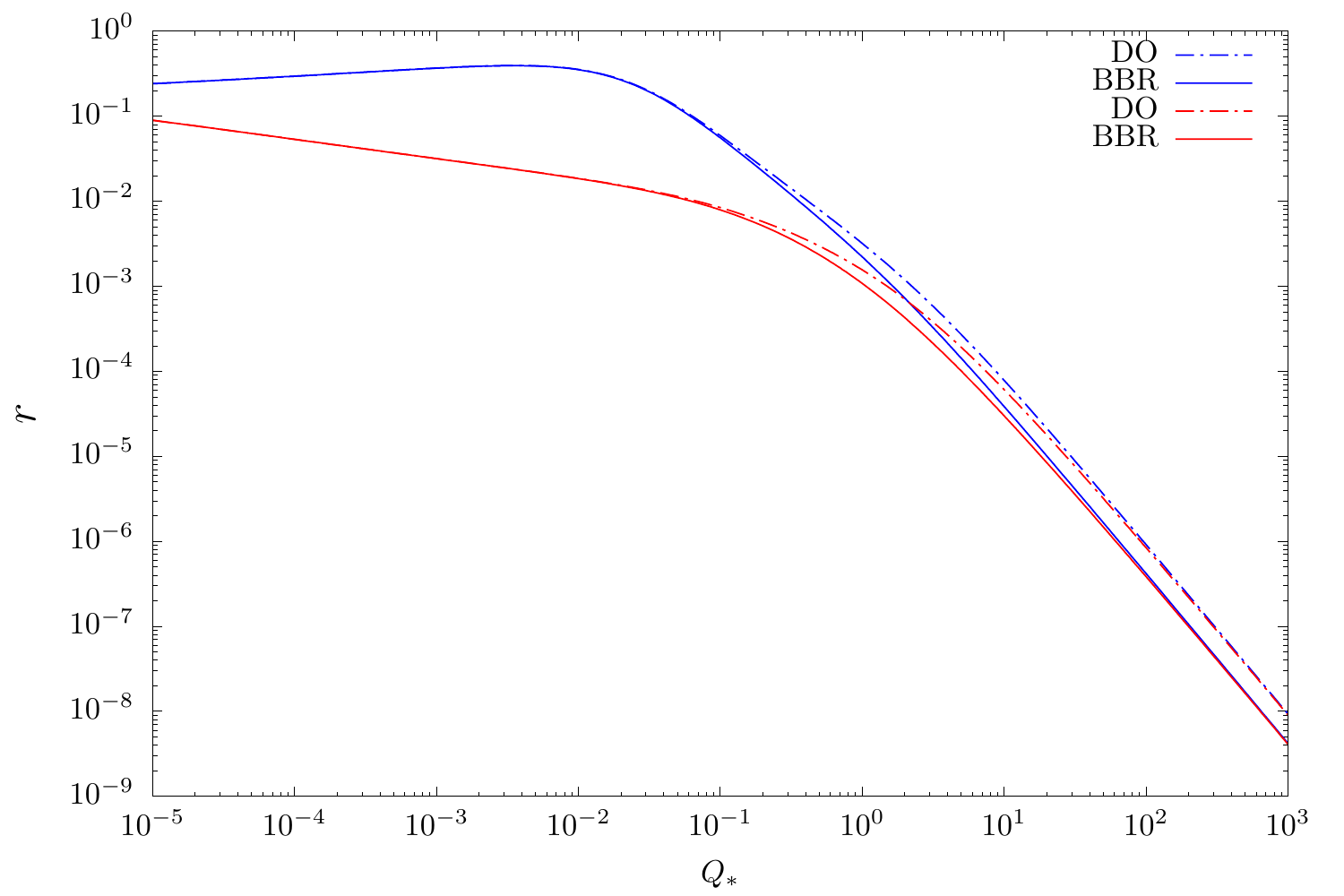}
    \caption{$\Upsilon \propto T^3/\phi^2$}
  \end{subfigure}  
  \caption{Predictions for a quadratic potential considering $60$ e-folds of 
expansion. Shown are at the top the spectral index and at the bottom the 
tensor-to-scalar ratio. Red lines denote $n_*=n_{BE}$ and blue lines $n_*=0$.}
  \label{fig::vp2}
  
\end{figure}

\begin{figure}[h!]
  \centering
  \begin{subfigure}[b]{0.47\textwidth}
    \includegraphics[width=\textwidth]{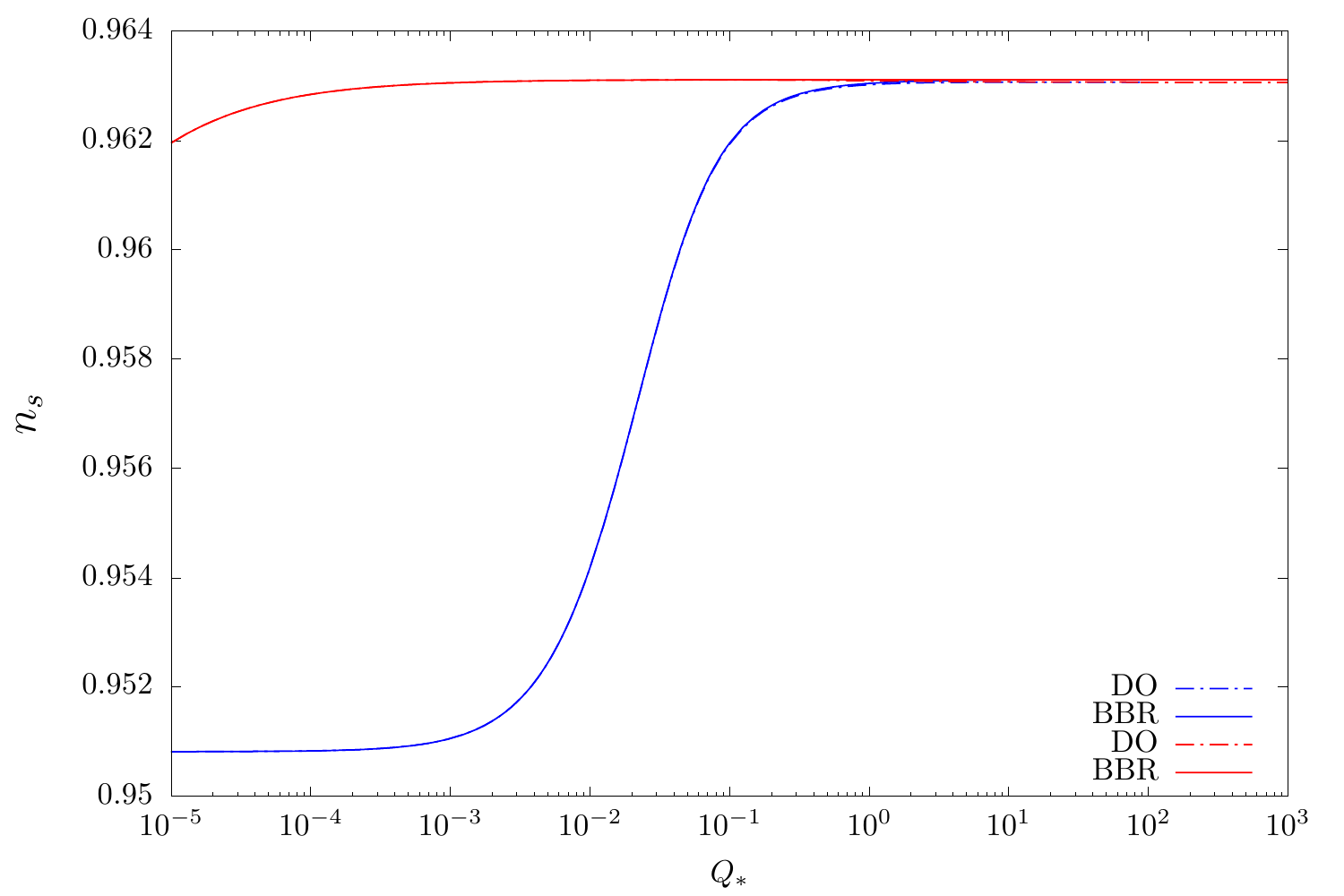}
    \caption{$\Upsilon \propto \phi^2$}
  \end{subfigure}
~
  \begin{subfigure}[b]{0.47\textwidth}
    \includegraphics[width=\textwidth]{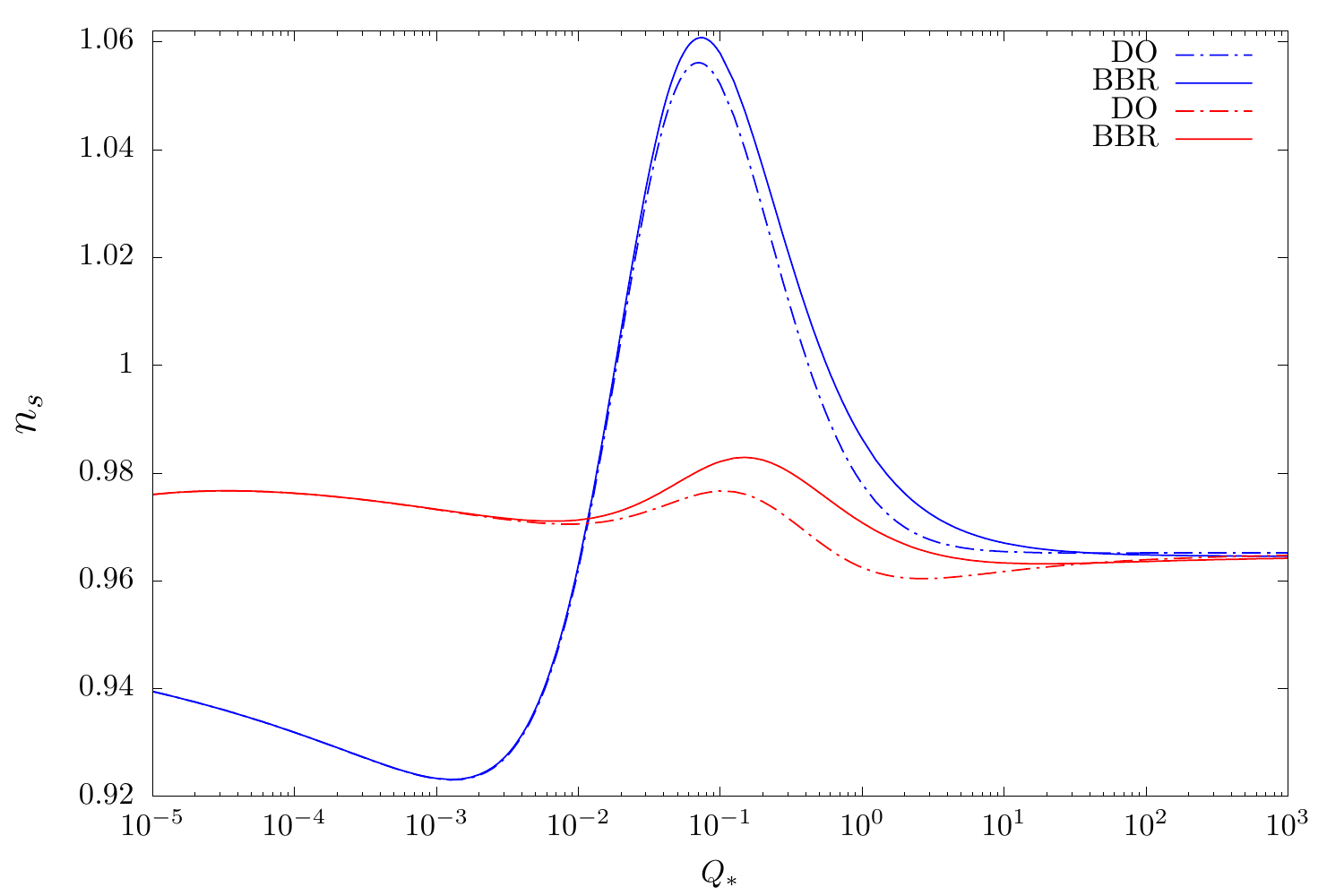}
    \caption{$\Upsilon \propto T^3/\phi^2$}
  \end{subfigure}\\
  \begin{subfigure}[b]{0.47\textwidth}
    \includegraphics[width=\textwidth]{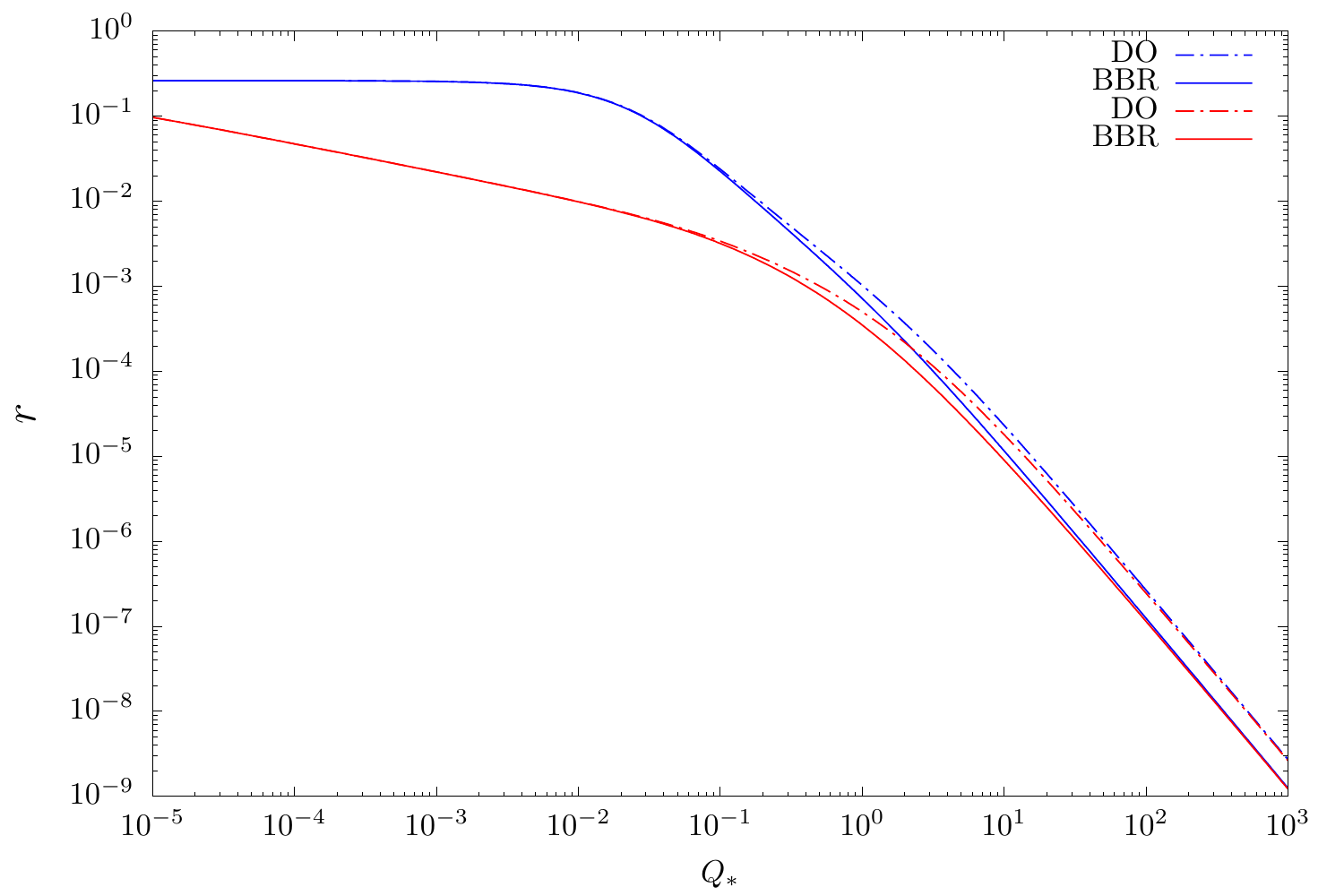}
    \caption{$\Upsilon \propto \phi^2$}
  \end{subfigure}
~
  \begin{subfigure}[b]{0.47\textwidth}
    \includegraphics[width=\textwidth]{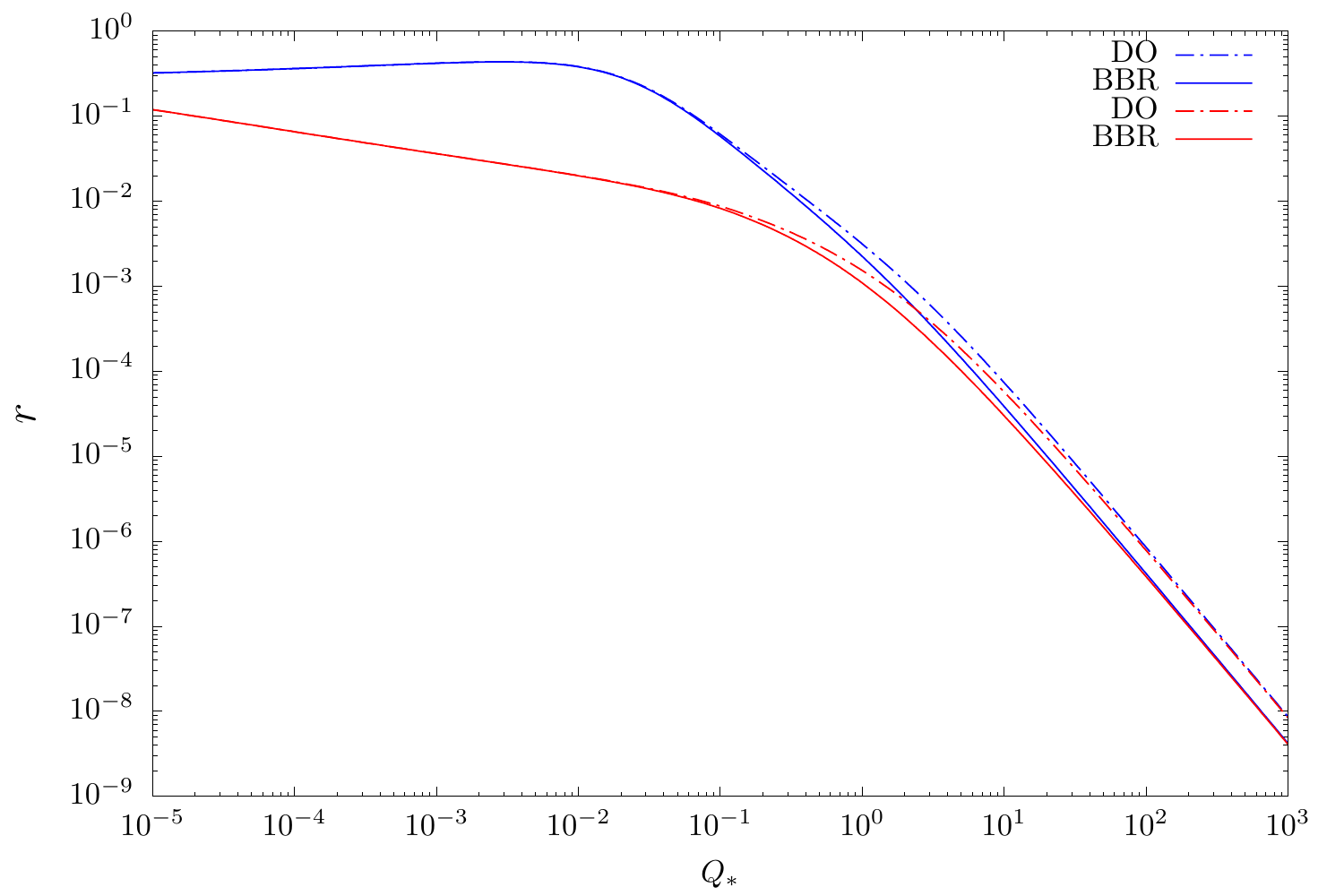}
    \caption{$\Upsilon \propto T^3/\phi^2$}
  \end{subfigure}  
  \caption{Predictions for a quartic potential considering $60$ e-folds of expansion. Shown are at the top the spectral index and at the bottom the 
tensor-to-scalar ratio. Red lines denote $n_*=n_{BE}$ and blue lines $n_*=0$.}
  \label{fig::vp4}
  
\end{figure}

Two different possibilities for $n_*$ are considered, the Bose-Einstein 
distribution $n_*=n_{BE}$ (red lines on the figures below) and 
$n_*=0$ (blue lines) , which, as will be seen, can lead to considerable 
differences in the weak dissipative regime. Three type of potentials are 
considered: quadratic, quartic and hybrid; and two types of dissipative 
coefficient: $\Upsilon \propto \phi^2$, i.e. quadratic in the field 
but $T-$independent (on the left of each figure), and a cubic $T-$ dependent 
one, $\Upsilon \propto T^3/\phi^2$, on the right. Note that for $T-$ dependent 
dissipative coefficients, the field spectrum expression \eqref{Deltaphi} 
does not hold in the strong dissipative regime when $Q \gtrsim \mathcal{O}(1)$. 
In that case, we have a coupled system of inflaton-radiation fluctuations, 
which gives rise to an enhancement (reduction) of the amplitude of the 
field spectrum for positive (negative) powers of 
$T$ \cite{Graham:2009bf, Bastero-Gil2011}. Given that this effect is model 
dependent, and again, to simplify the discussion, we will not take this 
into account. Notice however that this would only affect the results in the 
strong dissipative regime for the LHS figures. Our main aim is to compare 
results between the BBR and DO approaches.  For the spectral index in 
the strong dissipative regime they would give the same prediction, 
so adding the growing/decreasing mode to the calculation adds nothing 
to the present discussion. And for the tensor-to-scalar ratio, both 
will be similarly further suppressed, but keeping the ratio $r_{BBR}/r_{DO} \simeq 3/4$ when $Q \gg1$. Anyhow, Figs.~\ref{fig::vp2} and \ref{fig::vp4} show the predictions for a quadratic ($V(\phi) = m^2 \phi^2$)
and a quartic ($V(\phi) = \lambda \phi^4$) chaotic potential, respectively. As mentioned before, the  
quadratic dissipative coefficient are on the left, and the $T-$dependent 
coefficient on the right. As expected, noticeable differences only show up for 
the spectral index at $Q \sim 1$ in each case. In other dissipative regimes  
the differences are negligible. However, for the combination of a quartic 
chaotic potential and a quadratic and $T-$independent dissipative coefficient, 
there are no differences in the spectral index in any regime. For this 
potential we have that $\epsilon= 2 \sigma$, and therefore 
from \eqref{dlambda0} this gives $\theta=0$. 

\begin{figure}[h!]
  \centering
  \begin{subfigure}[b]{0.47\textwidth}
    \includegraphics[width=\textwidth]{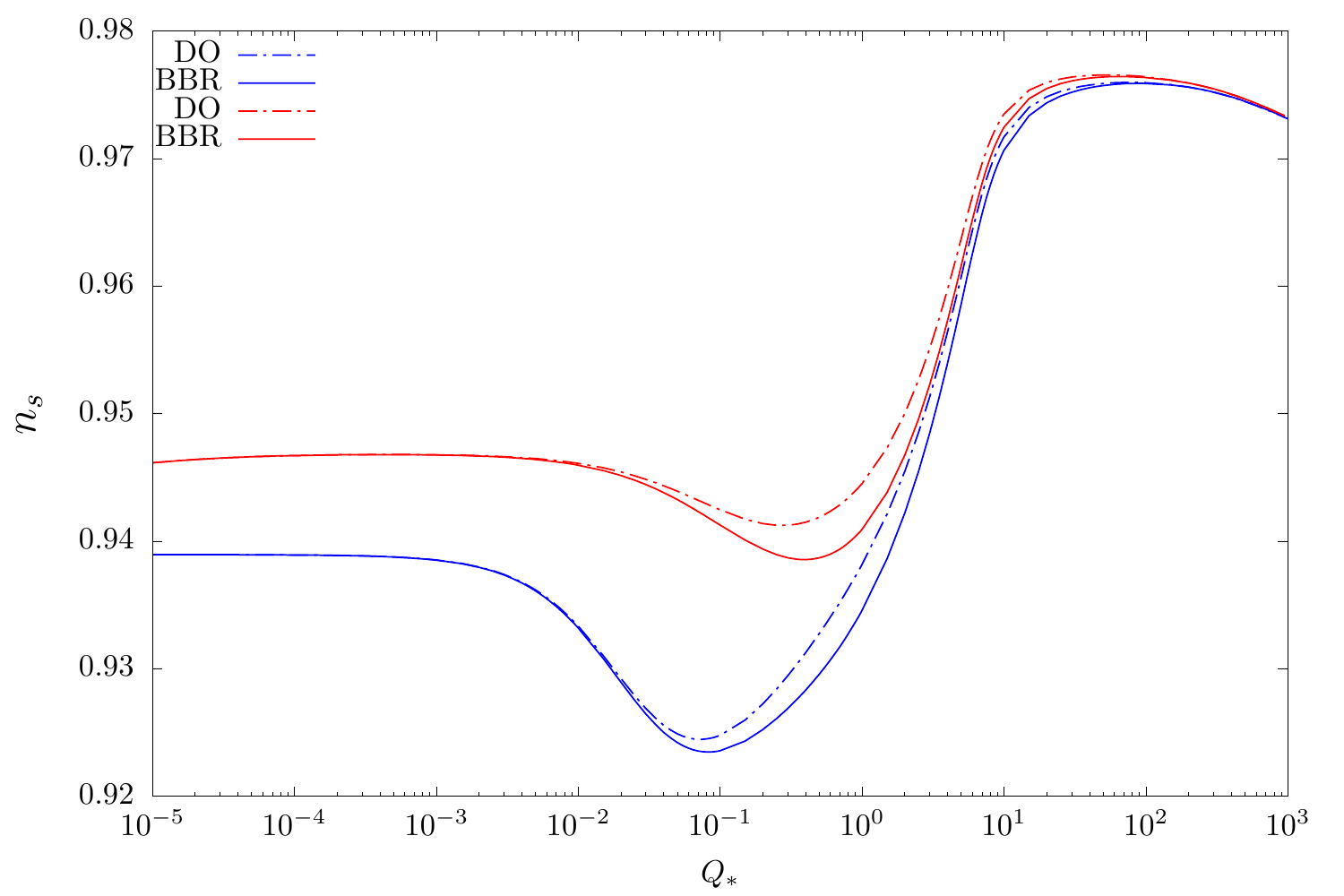}
    \caption{$\Upsilon \propto \phi^2$}
  \end{subfigure}
~
  \begin{subfigure}[b]{0.47\textwidth}
    \includegraphics[width=\textwidth]{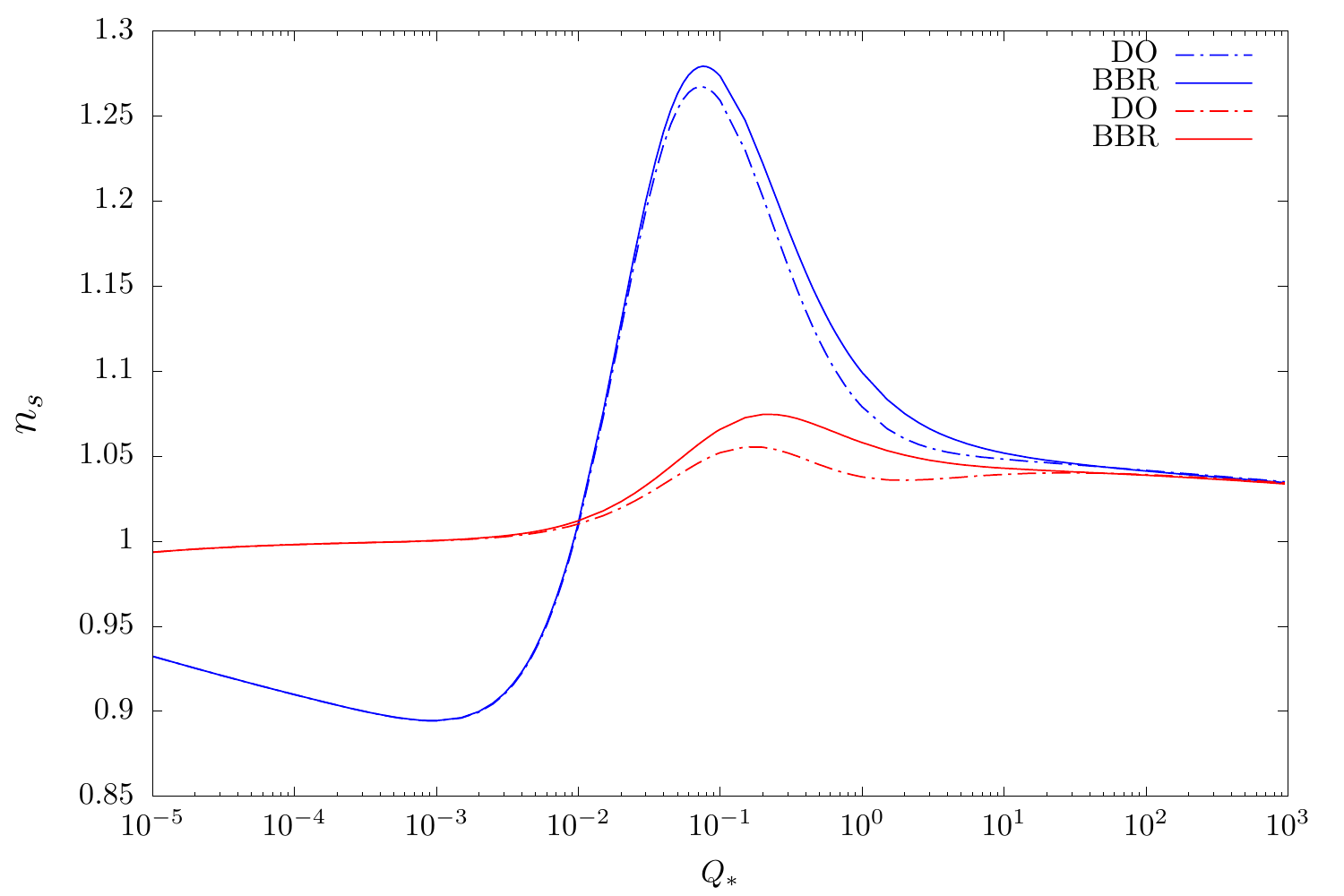}
    \caption{$\Upsilon \propto T^3/\phi^2$}
  \end{subfigure}\\
  \begin{subfigure}[b]{0.47\textwidth}
    \includegraphics[width=\textwidth]{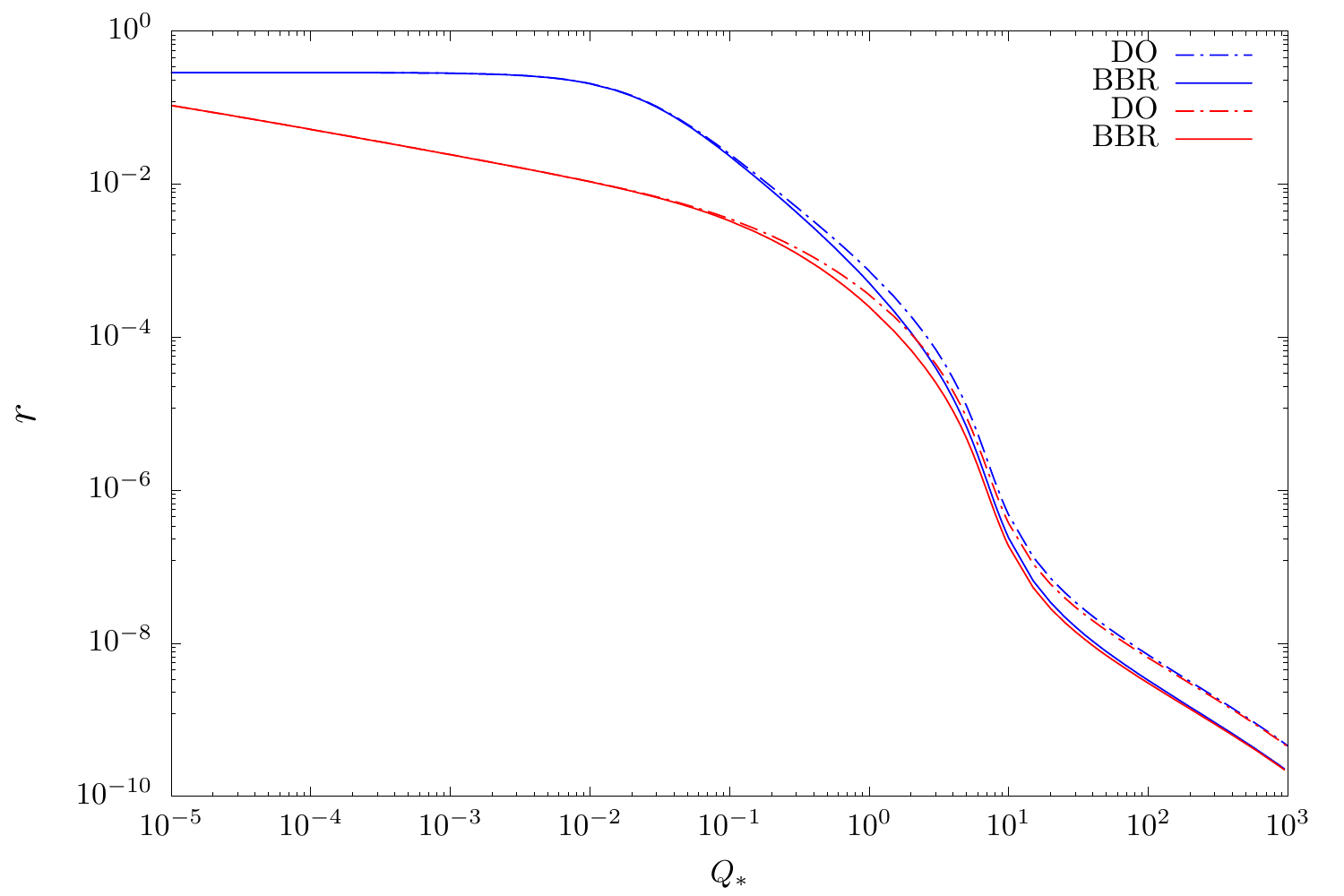}
    \caption{$\Upsilon \propto \phi^2$}
  \end{subfigure}
~
  \begin{subfigure}[b]{0.47\textwidth}
    \includegraphics[width=\textwidth]{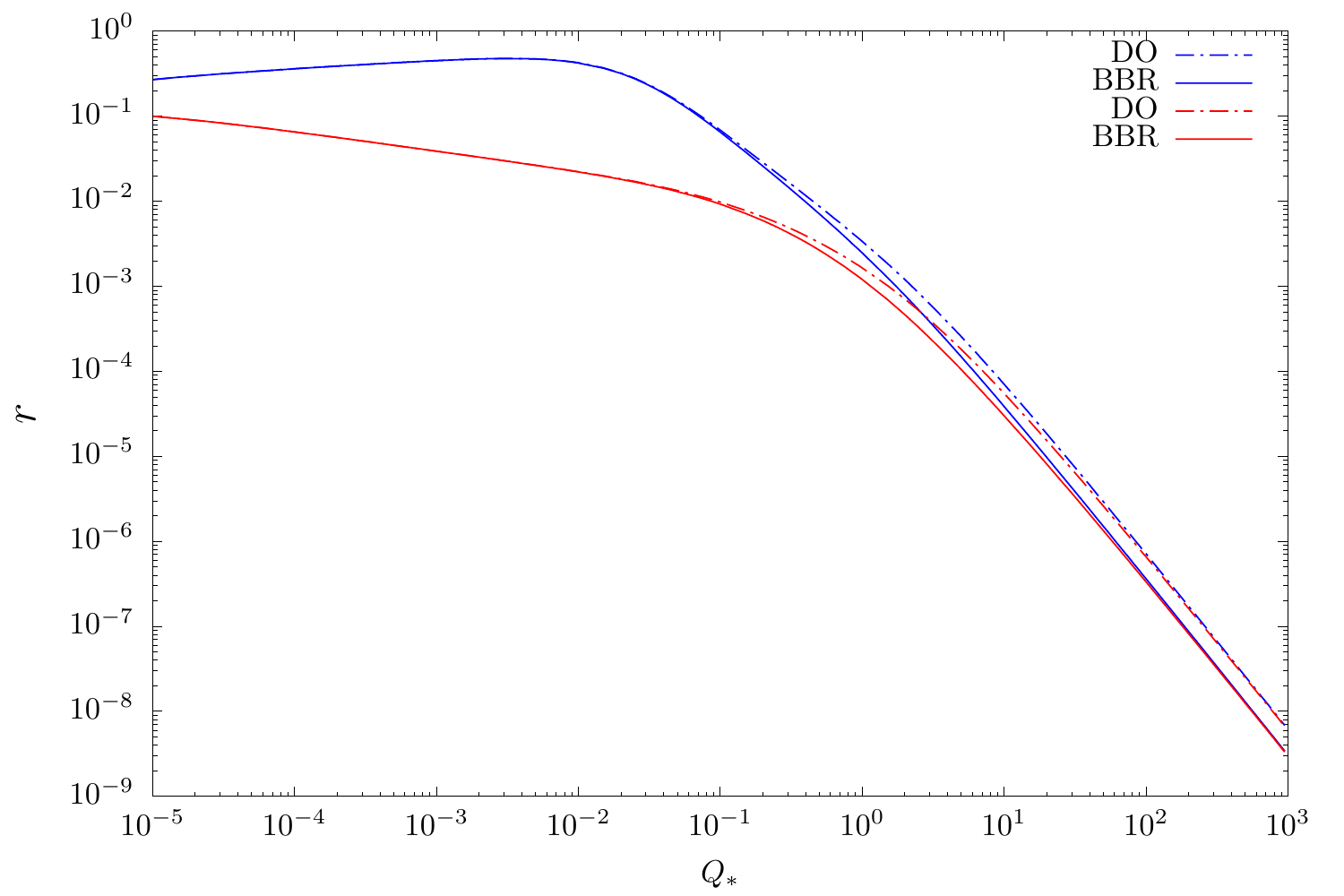}
    \caption{$\Upsilon \propto T^3/\phi^2$}
  \end{subfigure}  
  \caption{Predictions for a hybrid potential considering $60$ e-folds of 
expansion.  Shown are at the top the spectral index and at 
the bottom the tensor-to-scalar ratio. Red lines denote $n_*=n_{BE}$ and blue lines $n_*=0$.}
  \label{fig:vhyb}
  
\end{figure}

The hybrid potentials are also examined, with results shown in
Fig.~\ref{fig:vhyb}. This potential has two scalar fields,
although during slow-roll inflation only the inflaton degrees of
freedom are excited. Then, most of inflation takes place until $\phi$
gets to a critical value $\phi_c$, where the so-called waterfall field is
relevant for the background dynamics, bringing inflation to its end
soon after that. Therefore, during slow-roll, we can set the waterfall
field to zero and write the potential as
\begin{equation}
	V(\phi) = V_0 \left(1+\frac{k}{2} \phi^2 \right).
\end{equation} 
As a way to impose a condition on $k$, we required $\eta(\phi_c) = 0.1$. Since this does not fix all the required 
degrees of freedom, a condition on $\phi_c$ is also imposed. 
For quadratic dissipation, $\phi_c=0.1$ was taken, whereas for 
$\Upsilon \propto T^3/\phi^2$ it was $\phi_c=0.0045$, since it was 
complicated to get 60 e-folds of expansion with higher values. Anyway, the same qualitative behaviour was found, i.e., bigger differences in the spectral index when dissipation and expansion occur at similar rates.

\newpage
\section{Discussion}

In this article we have reexamined the two existing 
analytical approaches to compute
the comoving curvature perturbation and the observables derived from
it within the WI scenario. This is the first such explicit comparison 
between them,
perhaps because one would expect the results to agree due to a number of 
reasons including the gauge invariant nature of General Relativity. 
However, the formulas for the amplitude of the primordial 
spectrum differ, with a ``$Q/4$-difference'' spoiling the equivalence at 
leading order. This has consequences especially for the spectral index, with differences in the predicted values of order $\mathcal{O}(10^{-3})$ in dissipative regions where $Q \sim 1$; 
this was shown both analytically and in specific realisations. 

The origin of the discrepancies was found to be the different approximations 
performed in each approach. On the one hand, DO considers the time variation of 
the velocity field to be negligible, whereas BBR does so with the radiation 
momentum perturbation. Each assumption leads to different results because 
part of the dilution of the momentum perturbation is encompassed in the 
time variation of the velocity field. Given the dependence 
of $\mathcal{R}$ on $\Psi$, we conclude that the BBR approximation is a 
more sensible one, otherwise one would be underestimating the damping 
effect of expansion on the momentum perturbation, and consequently, 
on the curvature perturbation and the observables depending directly on 
that magnitude. Once this point is recognized, both methodologies are 
consistent at leading order. This is explained by the fact that both approaches considered the slow-roll approximation to be also valid for perturbed quantities. In Appendix \ref{s::aplt}, we explicitly show through numerical simulations that this is a well-founded assumption, in particular for the momentum perturbation. In this way, we are entitled to ignore the higher order time derivatives in the equations of motion of perturbations. Other terms that are consistently neglected at leading order are the metric and source perturbations, as well as the coupling between radiation and the metric perturbations. However, we do account for those terms in Appendix \ref{s::correctionsR}, where we follow the same program as before, obtaining an expression for the curvature perturbation valid at next-to-leading order. We found that there are no corrections to $\mathcal{R}$ at that order, since the total gauge invariant momentum perturbation and $\rho+p$ change by the same factor.


\appendix

\section{Slow-roll parameter $\theta$}
\label{s::tdc}
In order to derive the slow-roll parameter $\theta$ for a general $T$-dependent dissipative coefficient, $\Upsilon \propto T^c/\phi^m$, the evolution of 
the temperature in the slow-roll regime is needed.  Using the 
Stefan-Boltzmann law together with the slow-roll equation 
for $\rho_r$ gives
\begin{equation}
	\rho_r = C_r T^4 \simeq \frac{3}{4}Q\dot{\phi}^2 ,
\end{equation}
which leads to 
\begin{eqnarray}\label{dtn}
	\frac{d \ln T}{dN} & = & 
\frac{1}{2} (\epsilon - \eta) + \frac{1-Q}{4Q} \theta \,. 
\end{eqnarray}
Therefore,
\begin{eqnarray}
    \frac{d \ln \Upsilon}{dN} & = & c \frac{d \ln T}{dN} - m \frac{d \ln \phi}{dN}\\
    & = & c \left[\frac{1}{2}(\epsilon-\eta)+\frac{1-Q}{4Q}\theta\right] + m \sigma \,,
\end{eqnarray}
and thus, 
\begin{equation}
    \theta = \frac{Q}{1+Q} \frac{d \ln Q}{dN}=\frac{4Q}{4-c + (4+c)Q}\left[\left(1+\frac{c}{2}\right)\epsilon - \frac{c}{2}\eta + m\sigma \right] \,.
\label{dlambda}
\end{equation}
In particular for a monomial dissipative coefficient 
$\Upsilon \propto \phi^n$, setting $c=0$ and $m=-n$ 
in \eqref{dlambda} gives,
\be \label{dlambda0}
\theta = \frac{Q}{1+Q}(\epsilon - n\sigma) \,.
\ee

\section{Approach-Independent part of the Spectral Index}

In this section a derivation is outlined for an analytical expression for $\tilde{n}$,
\begin{equation}
  \tilde{n}  =  \frac{d}{dN}\ln \left[\left(\frac{\dot{\phi}}{2 m_P^2 \epsilon H}\right)^2 \Delta^2_{\delta\phi}\right] \,,
\end{equation}
which as introduced in \eqref{nsi}, is independent of the approach to compute the comoving curvature perturbation and its corresponding power spectrum. The inflaton power spectrum is given by:
\be
\Delta^2_{\delta\phi} =
\left(\frac{H}{2 \pi} \right)^2 (1 + 2 n_* + \frac{T}{H} F[Q])\,,
\ee
where
\be
F[Q]=\frac{12Q8^Q [\Gamma(3/2+3Q/2)]^3}{(1+3Q)\Gamma(1+3Q/2)\Gamma(5/2+3Q)} \,,
\ee
and either $1+ 2n_* \simeq 1$, or $1 + 2 n_*\simeq 2T/H$ for the Bose-Einstein distribution. Therefore, in general, $\tilde{n}$ is given by
\begin{equation}
 \tilde{n} =   2\eta -6 \epsilon +  \frac{d \ln (1 + n_*+ (T/H)F[Q])}{dN} \,, 
\end{equation}
where \eqref{eq::srdph} and \eqref{srw0} are used, 
together with \eqref{srwi1} in order to compute the derivative of the 
slow-roll parameter $\epsilon$. On the other hand, the last term on the 
RHS is model dependent. For instance, consider the case with $n_*=0$, so that
\bea
\frac{d \ln (1 + (T/H) F[Q])}{dN}
&=& \frac{(T/H) F[Q]}{1 + (T/H) F[Q]} \left[\frac{1}{4} ( 6 \epsilon -2 \eta)  + \left(\frac{1-Q}{4Q} +  \frac{1+Q}{Q} F^\prime[Q] \right)\theta \right] \,,
\eea
where we have defined $F^\prime[Q]= d \ln F[Q]/d \ln Q$, and we have 
used \eqref{dtn} and \eqref{dlambda} together with the 
definition of $\epsilon$. Collecting all the terms gives for $\tilde n$,
\be
\label{ntilde}
\tilde n  = \left(1 - \frac{(T/H) F[Q]}{1 + (T/H) F[Q]} \right) ( 2 \eta - 6 \epsilon) + \frac{(T/H) F[Q]}{1 + (T/H) F[Q]} \left( \frac{1-Q}{4Q} + \frac{1 +Q}{Q} F^\prime[Q]\right) \theta \,.
\ee
This expression can be simplified in the weak (WDR) and strong (SDR) 
dissipative limits. For the former, the dissipative ratio 
satisfies $Q \ll 1$, which gives
\be
\label{ntw}
\tilde n  \simeq \left(1 - \frac{\pi}{2}\frac{T}{H}Q  \right) ( 2 \eta - 6 \epsilon) + \frac{5 \pi}{2} \frac{T}{H} \theta \,,
\ee
whereas for strong dissipation
\begin{equation}\label{nts}
\tilde n  \simeq \left(\frac{3}{4} - \frac{1}{4} \frac{H/T}{\sqrt{3 \pi Q}}  \right) ( 2 \eta - 6 \epsilon) + \left(1- \frac{H/T}{\sqrt{3 \pi Q}}\right)  \frac{\theta}{4} \,. 
\end{equation}

A similar procedure can be followed for the case $n_* = n_{BE}$, 
where the expression for $\tilde n$ now reads
\be
\label{ntildeth}
\tilde n  = \frac{3}{4}( 2 \eta - 6 \epsilon) + \frac{1-Q}{4Q} \theta + \frac{F[Q]}{2 + F[Q]} F^\prime[Q] \frac{1+Q}{Q} \theta \,.
\ee
Taking again the limits for weak and strong dissipation, gives  
for $Q\ll 1$,
\be
\label{ntthw}
\tilde n  \simeq \frac{3}{4}( 2 \eta - 6 \epsilon) + \left(\frac{1}{4Q}+ \pi -\frac{1}{4} \right) \theta \,,
\ee
whereas for $Q \gg 1$ it reads,
\be
\label{ntths}
\tilde n  \simeq \frac{3}{4}( 2 \eta - 6 \epsilon) + \left( \frac{1}{4}-\frac{1}{\sqrt{3 \pi Q}}  \right) \theta \,.
\ee

\section{Corrections to the comoving curvature perturbation at linear order}\label{s::correctionsR}
In this section, we intend to derive a formula for the comoving curvature perturbation, including corrections at higher order, if any. In doing so, some gaps  in the derivation obtained at zeroth order in Section \ref{BBR} are filled. There are two central parts in the derivation, which we can identify by looking at the definition of the comoving curvature perturbation, 
\begin{equation}\label{apc_r}
	\mathcal{R}=-\frac{H}{\rho+p} \Psi_{T}^{G I}=-\frac{H}{\rho+p}\left(\Psi_{\phi}^{G I}+\Psi_{r}^{G I}\right).
\end{equation}
Clearly, both the numerator (through $\Psi_r^{GI}$) and the denominator (through $\rho_r$) require ``improved'' expressions in order to get one for $\mathcal{R}$. To get such formulas, the plan in both cases is to solve perturbatively the corresponding equations, using the slow-roll approximation when appropriate. 

\subsection{Corrections to $\rho_r$}

First, we will compute the corrections for the denominator of \eqref{apc_r}, given by
\begin{equation}\label{rpp}
	\rho + p = \dot{\phi}^2 + \frac{4}{3} \rho_r.
\end{equation}
Since the higher order terms come from the radiation energy density, henceforth we will focus on this variable. At zeroth order, it satisfies
\begin{equation}
	\rho_r^{(0)} \simeq \frac{3}{4} Q \dot{\phi}^2,
\end{equation}
which no longer holds at the desired level of approximation. To solve this, we posit that it can be written as a series in the slow-roll parameter $\epsilon$ like
\begin{equation}
	\rho_r = c_0 + c_1 \epsilon + \mathcal{O}(\epsilon^2).
\end{equation}
Naturally, the first term corresponds to the usual slow-roll approximation, i.e., $\rho_r^{(0)} = c_0$. The second coefficient can be found by replacing this into the equation of motion of the radiation energy density \eqref{eq::radd1}, such that
\begin{equation}
	\dot{c_0} + \dot{c_1}\epsilon + c_1 \dot{\epsilon} + 4H(c_0 + c_1 \epsilon) = \Upsilon \dot{\phi}^2 = 4Hc_0,
\end{equation}
where the second and third term on the LHS clearly introduce corrections at higher order. In consequence, the second coefficient is given by
\begin{equation}
	c_1 = -\frac{\dot{c_0}}{H\epsilon} = - \frac{\rho_r^{(0)}}{4 \epsilon} \left(2(\epsilon - \eta) + \frac{1-Q}{Q} \theta \right),
\end{equation}
so the radiation energy density now reads
\begin{equation}
	\rho_r \simeq \rho_r^{(1)} = \frac{3}{4}Q\dot{\phi}^2 \left(1 - \frac{\epsilon - \eta}{2} - \frac{1-Q}{4Q} \theta \right).
\end{equation}
As such, \eqref{rpp} becomes 
\begin{equation}\label{rppf}
	\rho + p \simeq \dot{\phi}^2 \left\{1+Q\left(1-\frac{\epsilon - \eta}{2} - \frac{1-Q}{4Q} \theta \right)\right\} \;.
\end{equation}

\subsection{Corrections to $\Psi_r$}

In order to simplify the notation, we drop the superindex ``GI'' from the perturbations, but all of them should be understood to be in their gauge invariant form. Having said that, we write $\Psi_r$ in terms of its contributions at each order of approximation, i.e.,
\begin{equation}
	\Psi_r \simeq \Psi_r^{(0)} + \Psi_r^{(1)},
\end{equation}
where $\Psi_r^{(0)} = Q\Psi_{\phi}$, as shown in Section \ref{BBR}. In this way, the equation of motion of the momentum perturbation \eqref{eq::dq} becomes
\begin{equation}
	\dot{\Psi}_{r}^{(0)}+3 H\left(\Psi_{r}^{(0)}+\Psi_{r}^{(1)}\right) \simeq \Upsilon \Psi_{\phi}-\frac{1}{3} \rho_{r}-\frac{4}{3} \rho_{r} \mathcal{A},
\end{equation}
where we have omitted $\dot{\Psi}_{r}^{(1)}$, as it is a higher order term. Then, the correction to the radiation momentum perturbation is given by
\begin{equation}\label{ho-pr}
	\Psi_{r}^{(1)}=-\frac{1}{9 H}\left[\delta \rho_{r}+4 \rho_{r} \mathcal{A}\right]-\frac{\dot{\Psi}_{r}^{(0)}}{3 H}.
\end{equation}
In order to continue, we need to invoke the other equations of motion, in particular \eqref{eq::dp} and \eqref{eq::dr}. In both cases, we will use the slow-roll approximation, neglecting the higher time derivatives of each variable, which leads to the following equations 
\begin{equation}
	3 H(1+Q) \dot{\delta\phi}+V_{\phi \phi} \delta \phi \simeq-\dot{\phi} \delta \Upsilon+\Upsilon \dot{\phi} \mathcal{A}+6 H \dot{\phi} \mathcal{A},
\end{equation}
\begin{equation}\label{eq::srho}
	4H \delta \rho_r \simeq \delta \mathcal{Q}_r + \mathcal{Q}_r \mathcal{A},
\end{equation}
where we have also taken the $k \ll aH$ limit. From the first equation and the definition of the slow-roll parameter $\eta$, we can get the useful relation 
\begin{equation}\label{eq::axdphi}
	\frac{\dot{\delta \phi}}{\dot{\phi}} \simeq-\frac{H}{\dot{\phi}} \eta \delta \phi-\frac{Q}{1+Q} \frac{\delta \Upsilon}{\Upsilon}+\frac{2+Q}{1+Q} \mathcal{A}.
\end{equation}

Next, we will concentrate on getting an expression for $\delta \rho_r$. With this goal in mind, take the definition of the source, $\mathcal{Q}_r = \Upsilon \dot{\phi}^2$, such that its perturbation reads
\begin{equation}\label{eq::dQr}
	\frac{\delta \mathcal{Q}_{r}}{\mathcal{Q}_{r}}=\frac{\delta \Upsilon}{\Upsilon}+2\left(\frac{\dot{\delta\phi}}{\dot{\phi}}-\mathcal{A}\right).
\end{equation}
Furthermore, since $\Upsilon \propto T^c/\phi^m$ and $\rho_r \propto T^4$, we have that
\begin{equation}\label{eq::dUu}
	\frac{\delta \Upsilon}{\Upsilon} = c \frac{\delta T}{T} - m \frac{\delta \phi}{\phi} = \frac{c}{4} \frac{\delta \rho_r}{\rho_r} - m \frac{\delta \phi}{\phi}.
\end{equation}
Then, plugging \eqref{eq::dQr} into \eqref{eq::srho}, and using \eqref{eq::axdphi} and \eqref{eq::dUu}, we get
\begin{equation}\label{dRh}
	\frac{\delta \rho_{r}}{9 H}=\frac{1+Q}{4-c+(4+c) Q} \frac{\dot{\phi}^2}{3H}Q\left[\frac{3+Q}{1+Q} \mathcal{A}-m \frac{1-Q}{1+Q} \frac{\delta \phi}{\phi}-2 \frac{H}{\dot{\phi}} \eta \delta \phi\right].
\end{equation}
\\

Turning our attention back to \eqref{ho-pr}, we need to compute the time derivative of the radiation momentum perturbation at zeroth order. Clearly, this is given by
\begin{equation}\label{dPr}
	\dot{\Psi}_{r}^{(0)}=\left(\dot{Q} \Psi_{\phi}+Q \dot{\Psi}_{\phi}\right),
\end{equation}
where 
\begin{equation}\label{dQl}
	\dot{Q} = H(1+Q)\theta,
\end{equation}
\begin{equation}
	\dot{\Psi}_{\phi}=-(\ddot{\phi} \delta \phi+\dot{\phi} \dot{\delta \phi}).
\end{equation}
We can conveniently write $\ddot{\phi}$ in terms of the slow-roll parameters through \eqref{eq::srdph}, which yields
\begin{equation}
	\ddot{\phi} \simeq H(\epsilon - \eta - \theta)\dot{\phi}.
\end{equation}
Then, invoking \eqref{eq::axdphi} once again, the time derivative of the field momentum perturbation reads
\begin{equation}
	\dot{\Psi}_{\phi} = 	H(\epsilon-2 \eta-\theta) \Psi_{\phi}+\dot{\phi}^{2}\left\{\frac{Q}{1+Q}\left[\frac{c}{4} \frac{\delta \rho_{r}}{\rho_{r}}-m \frac{\delta \phi}{\phi}\right]-\frac{2+Q}{1+Q} \mathcal{A}\right\},
\end{equation}
and, consequently, \eqref{dPr} becomes
\begin{equation}
	\dot{\Psi}_{r}^{(0)}=H \theta \Psi_{\phi}+Q\left[H(\epsilon-2 \eta) \Psi_{\phi}+\dot{\phi}^{2}\left\{\frac{Q}{1+Q}\left[\frac{c}{4} \frac{\delta \rho_{r}}{\rho_{r}}-m \frac{\delta \phi}{\phi}\right]-\frac{2+Q}{1+Q} \mathcal{A}\right\}\right],
\end{equation}
where, in addition, we have used \eqref{dQl}. Hence, plugging this into \eqref{ho-pr} and rearranging similar terms, we get
\begin{equation}
	\Psi_r^{(1)} = -\frac{1+(1+c)Q}{1+Q}\frac{\delta\rho_r}{9H} + \frac{\dot{\phi}^2}{3H} \frac{Q}{1+Q}\mathcal{A} - \frac{Q}{3}\left(\epsilon - 2\eta + \frac{\theta}{Q}\right) \Psi_{\phi} + \frac{\dot{\phi}^2}{3H}\frac{mQ^2}{1+Q}\frac{\delta\phi}{\phi}.
\end{equation}
Replacing \eqref{dRh} above, it follows that
\begin{eqnarray}\label{dprc}
	\Psi_r^{(1)} & = &  \frac{\dot{\phi}^{2}}{3 H} \frac{mQ}{1+Q} \frac{\delta \phi}{\phi}\left[Q + \frac{1+(1+c) Q}{4-c+(4+c) Q} (1-Q) \right] + \frac{2}{3}Q \eta \Psi_{\phi} \left[1 - \frac{1+(1+c) Q}{4-c+(4+c) Q} \right] \nonumber \\
	&&+\frac{\dot{\phi}^{2}}{3 H} \frac{Q}{1+Q} \mathcal{A} \left[1 - \frac{1+(1+c) Q}{4-c+(4+c) Q} (3+Q) \right] - \left[Q \frac{\epsilon}{3} + \frac{\theta}{3} \right] \Psi_{\phi} \;.
\end{eqnarray}
The remaining task is to write the terms proportional to $\delta \phi/\phi$ and $\mathcal{A}$ as functions of $\Psi_{\phi}$. The former can be easily done by considering the slow-roll parameter $\sigma$ together with \eqref{dlambda}, which yields
\begin{equation}\label{sigf}
	\sigma = - \frac{d \ln \phi}{dN} = - \frac{\dot{\phi}}{H\phi} =\frac{1}{m}\left[-\left(1+\frac{c}{2}\right) \epsilon+\frac{c}{2} \eta+\frac{4-c+(4+c) Q}{4 Q} \theta\right].
\end{equation}
Finally, the term proportional to $\mathcal{A}$ can be easily dealt with by noticing that
\begin{equation}
	\mathcal{A} = \epsilon \mathcal{R} = \epsilon (\mathcal{R}^{(0)} + \mathcal{R}^{(1)}),
\end{equation}
where, as usual, the second term on the RHS introduces higher order corrections (and which we are trying to compute). Thus, at this stage, it will be enough to keep the first term, such that
\begin{equation}\label{Af}
	\mathcal{A}^{(0)} = \epsilon \mathcal{R}^{(0)} = -\frac{\epsilon H}{\dot{\phi}^2}\Psi_{\phi}.
\end{equation}
Therefore, replacing \eqref{sigf} and \eqref{Af} in \eqref{dprc}, and after some algebra, we have got that
\begin{equation}
	\Psi_r^{(1)} = Q \left(-\frac{\epsilon}{2} + \frac{\eta}{2} - \frac{1-Q}{4Q}\theta\right)\Psi_{\phi},
\end{equation}
and, thus, the radiation momentum perturbation at next-to-leading order is 
\begin{equation}
	\Psi_r = Q \left(1 -\frac{\epsilon}{2} + \frac{\eta}{2} - \frac{1-Q}{4Q}\theta\right)\Psi_{\phi}.
\end{equation}
Notice that the total momentum perturbation now reads
\begin{equation}
	\Psi_T = \left\{1+Q\left(1-\frac{\epsilon-\eta}{2}-\frac{1-Q}{4 Q} \theta\right)\right\} \Psi_{\phi},
\end{equation}
i.e., it has the same correction factor as the one we found for $\rho + p$. In consequence, they cancel each other out in \eqref{apc_r}, so that
\begin{equation}
	\mathcal{R} = - \frac{H}{\dot{\phi}^2}\left[1 + \mathcal{O}(\epsilon^2)\right] \Psi_{\phi}.
\end{equation}

\section{Ratio between $\dot{\Psi}_r$ and $H \Psi_r$}\label{s::aplt}

\begin{figure}[h!]
  \centering
  \begin{subfigure}[b]{0.48\textwidth}
    \includegraphics[width=1.2\textwidth]{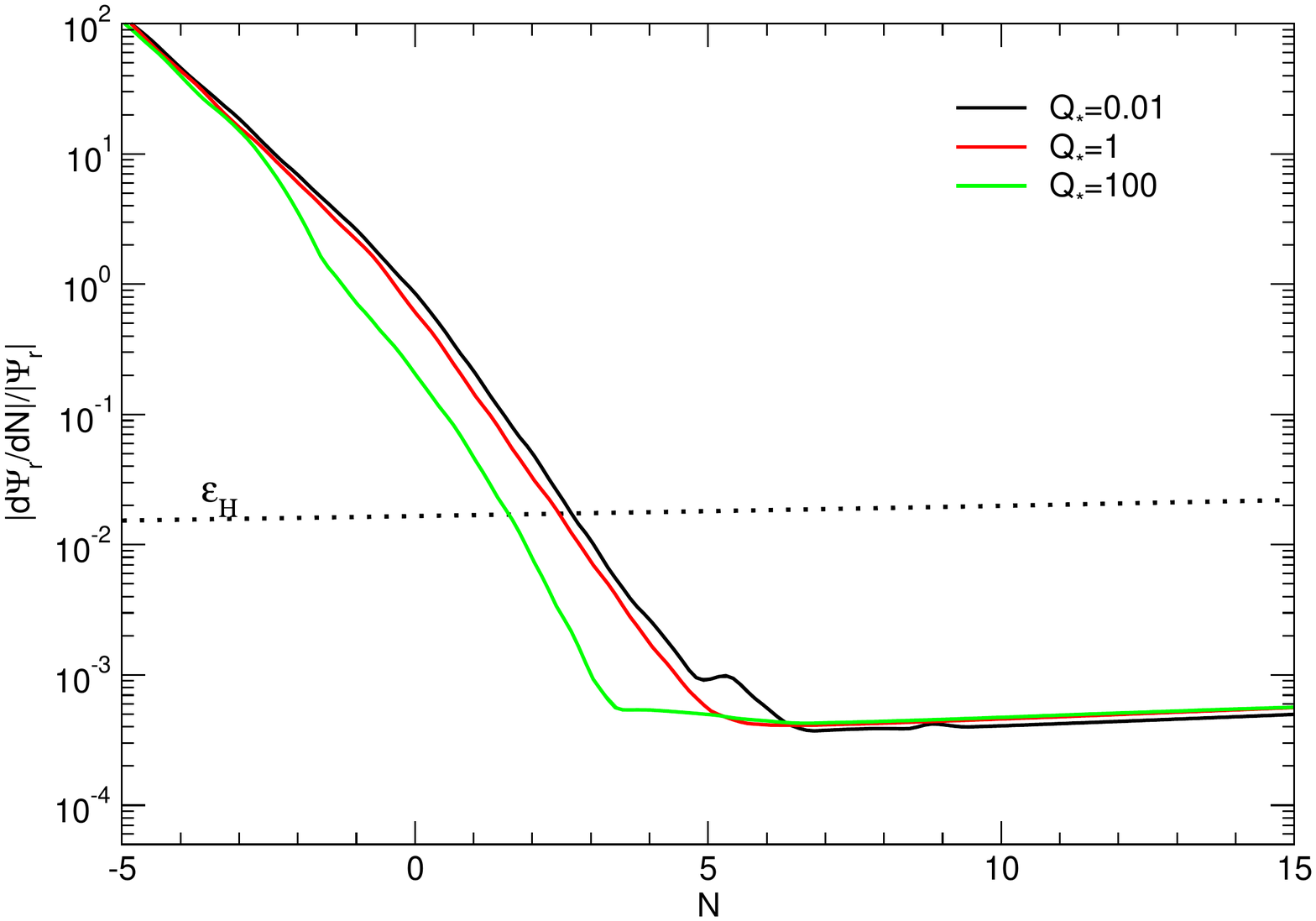}
    \caption{$\Upsilon \propto \phi^2$}
  \end{subfigure}
  ~
  \begin{subfigure}[b]{0.48\textwidth}
    \includegraphics[width=1.2\textwidth]{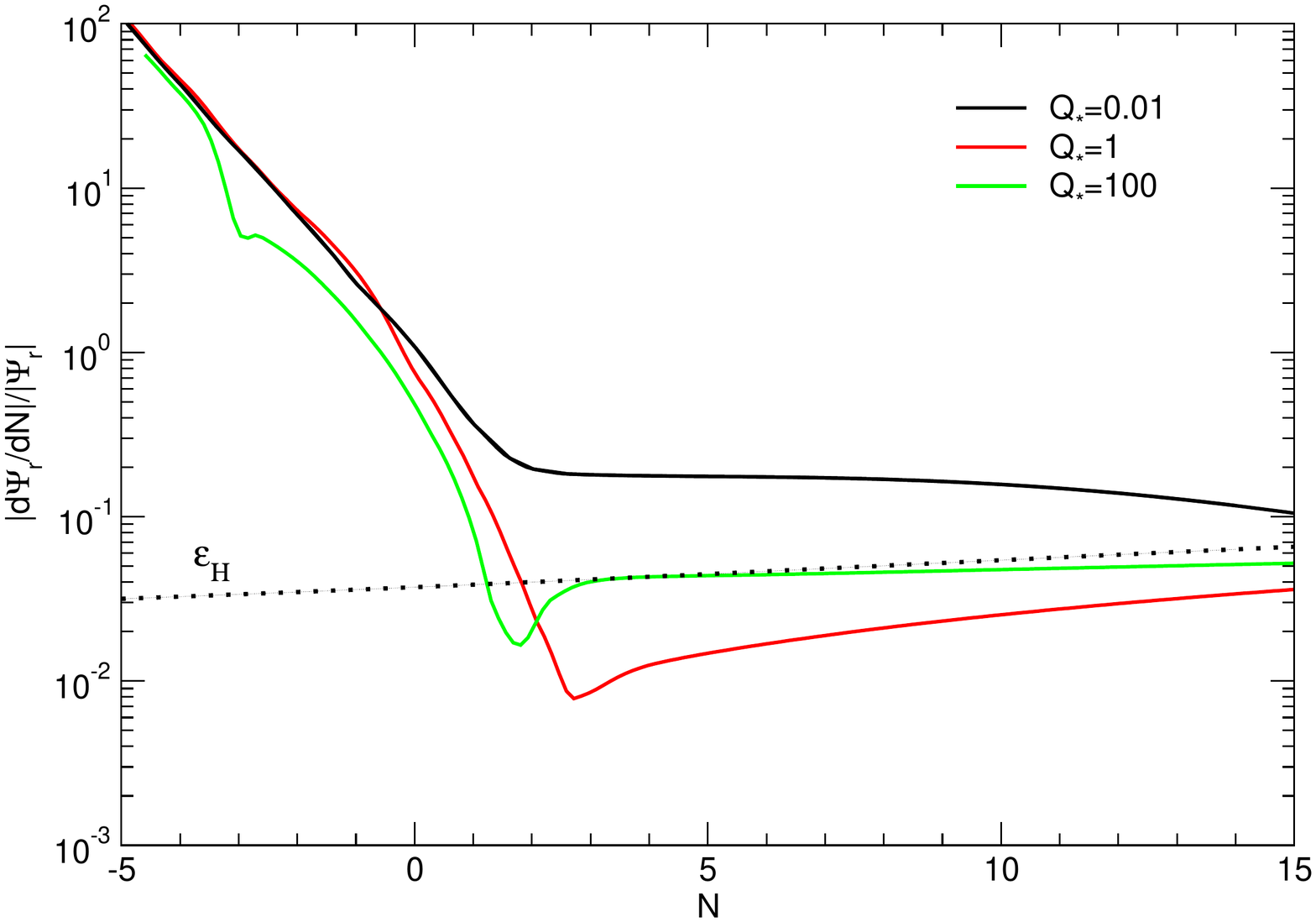}
    \caption{$\Upsilon \propto T^3/\phi^2$}
  \end{subfigure}
\caption{Ratio between $d \Psi_r/dN$ and $\Psi_r$ or, equivalently, $\dot{\Psi}_r$ and $H\Psi_r$, for a quartic potential. The dotted line shows the evolution of the slow-roll parameter $\epsilon \equiv \epsilon_H$ with $N$. The figures are plotted such that $k=aH$ at $N=0$, and the end of inflation happens at $N=60$.  }
  \label{fig:apdx}
  
\end{figure}

As a way of expanding on the argument about the discrepancies between DO and BBR, and the validity of the slow-roll approximation for perturbed variables, we present in Fig.~\ref{fig:apdx} the result of numerical simulations showing the ratio between $d \Psi_r/dN$ and $\Psi_r$ for a quartic potential and two types of dissipative coefficients, $\Upsilon \propto \phi^2$ (left) and $\Upsilon \propto T^3/\phi^2$ (right). The slow-roll parameter $\epsilon \equiv \epsilon_H$ is shown as a reference. Thus, in the first case, it can be seen that the ratio is of order $\mathcal{O}(\epsilon^2)$, whereas for the $T-$dependent one, it is of order $\mathcal{O}(\epsilon)$. In either case, at least for a leading order approximation, $\dot{\Psi}_r$ can be safely assumed to be negligible in comparison to $H \Psi_r$. In this way, the time derivative of the velocity field, given by
\begin{equation}
	\dot{v}=\frac{3}{4} \frac{k}{a \rho_{r}}\left(\dot{\Psi}_{r}-\frac{\dot{\rho}_{r}}{\rho_{r}} \Psi_{r}-H \Psi_{r}\right),
\end{equation}
can be well approximated by
\begin{equation}
	\dot{v} \simeq -\frac{3}{4}\frac{k}{a} \frac{H}{\rho_r}\Psi_r.
\end{equation}
However, it is worth emphasizing that this is not negligible at a leading-order approximation, as discussed in Section~\ref{ss:dis}, because it encompasses part of the dilution of the radiation momentum perturbation. 

\acknowledgments
MBG is partially supported by MINECO grant FIS2016-78198-P, and Junta 
de Andaluc\'ia Project FQM-101. AB is supported by STFC. JRC is supported by the Secretary of Higher Education, Science, Technology and Innovation of Ecuador (SENESCYT).

\bibliographystyle{jhep}
\bibliography{References}

\end{document}